\documentclass[a4paper,11pt]{article}
\usepackage{amsfonts}
\usepackage[latin1]{inputenc}
\usepackage[T1]{fontenc}
\usepackage[english]{babel}
\usepackage{amssymb}
\usepackage{latexsym}
\usepackage{amsmath}
\usepackage{amsthm}
\usepackage{color}
\usepackage{graphicx}
\newcommand{\bbeta}{{\boldsymbol{\beta}}}

\newtheorem{thm}{Theorem}[section]

\newtheorem{lem}[thm]{Lemma}

\title{The adaptive Gril estimator with a diverging number of parameters}
\author{Mohammed El Anbari and Abdallah Mkhadri
\vspace{0.1cm} \\
Department of Mathematics, Faculty of Sciences Semlalia,\\
Cadi Ayyad University, B.P. 2390 Marrakesh, Morocco \\and Dept. de
math\'{e}matiques Bat\^{i}ment 425\\ Universit\'{e} Paris-Sud, 91405 Orsay
Cedex.}
\begin{document}
 \maketitle
\begin{abstract}
 We consider the problem of variables selection and estimation in linear regression
model in situations where the number of parameters diverges with the
sample size. We propose the adaptive Generalized Ridge-Lasso
(\mbox{AdaGril}) which is an extension of the the adaptive Elastic Net.
AdaGril incorporates information redundancy among correlated
variables for model selection and estimation. It combines the
strengths of the quadratic regularization and the adaptively weighted
Lasso shrinkage. In this paper, we highlight the grouped selection property for AdaCnet method
(one type of AdaGril) in the equal correlation case. Under weak conditions, we establish the oracle
property of AdaGril which ensures the optimal large performance when
the dimension is high. Consequently, it achieves both goals of
handling the problem of collinearity in high dimension and enjoys
the oracle property. Moreover, we show that AdaGril estimator achieves a Sparsity Inequality, i. e.,
a bound in terms of the number of non-zero components of the 'true' regression coefficient. This
bound is obtained under a similar weak Restricted Eigenvalue (RE) condition used for Lasso.
Simulations studies show that some particular cases of AdaGril outperform its competitors.
\end{abstract}
\section{Introduction}
 We consider the problem of variable selection and estimation for
general linear regression model
\begin{equation} \label{equ1}
\mathbf{y}=\mathbf{X}\boldsymbol{\bbeta}^{*}+ \boldsymbol{\varepsilon},
\end{equation}
where $\mathbf{y} = (y_{1}, ..., y_n)^t$ is an $n$-vector of
responses, $\mathbf{X} = (\mathbf{x}_{1}, ..., \mathbf{x}_{p})$ is a
$n$x$p$ design matrix of $p$ predictor vectors of dimension $n$,
$\boldsymbol{\bbeta}^{*}$ is a $p$-vector of unknown parameters which are to be
estimated, t stands for the transpose and $\boldsymbol{\varepsilon}$ is a
$n$-vector of (i.i.d.) random errors with mean $0$ and variance
$\sigma^2$. Without loss of generality we assume that the data are
centered.

When $p$ is large, selection of a small number of predictors that
contribute to the response leads often to a parsimonious model. It
amounts to assuming that $\boldsymbol{\bbeta}^{*}$ is sparse in the sense $s<p$ components are non-zero.
Denote the set of non-zero values by ${\cal A}=\{ j; | \bbeta_j^{*}| \neq 0 \}$. In this setting,
variable selection can improve on both estimation accuracy and
interpretation. Our goal is to determine the set $\cal{A}$ and to estimate the true corresponding coefficients.

Sparsity is associated to high dimensional data, where the number of
predictors $p$ is typically comparable or exceeds the sample size
$n$. The problem occurs frequently in genomics and preteomics
studies, functional MRI, tumor classification and signal processing
(cf. Fan and Li 2008). In many of these applications, we would like
to achieve both variable reduction and prediction accuracy.

Variable selection for high dimensional data has received a lot of
attention recently. In the last decade interest has focused on penalized regression methods
which implement both variable selection and coefficient estimation in a single procedure.
The most well known of these procedures are Lasso (Tishirani 1996,
Chen et al. 1998) and SCAD (Fan and Li 2001), which have good
computational and statistical properties.

In fact, there has been a large rapidly growing body of literature
for the Lasso and SCAD studies over the past few years. Osborne et
al. (2000) derived the optimality conditions associated with the
Lasso solution. Some theoretical statistical aspects of the
 Lasso estimator of the regression coefficients have been derived by
Knight and Fu (2000) in finite dimension setting. Many other
extensions for asymptotic and non asymptotic results can be found in
Zhang and Yu (2006) and
 Bunea et al (2007), etc.

Various extensions and modifications of the Lasso have been proposed
to ensure that on one hand, the variable selection process is
consistent and on the other hand, the estimated regression
coefficient has a fast rate of convergence. Fan and Li (2001)
 showed that the SCAD enjoys the oracle property, that is, the SCAD estimator
 can perform as well as the oracle if the penalization parameter is appropriately chosen. Fan and Peng (2004)
studied the asymptotic behavior of SCAD when the dimensionality of
the parameter diverges. Fan and Li (2001) showed that asymptotically
the Lasso estimates produce non-ignorable bias. Zou (2006) showed
that the Lasso has not the oracle property in finite parameter
setting as conjectured in Fan and Li (2001). Zhao and Yu (2008)
established the same result for $p>n$ case.

To overcome the bias problem of Lasso, Zou (2006) proposed the
adaptive Lasso estimator (AdaLasso) defined by
\begin{equation}\label{equ2}
{\hat{\boldsymbol{\bbeta}}}_{\mbox{AdaLasso}}=\arg\min_{\boldsymbol{\bbeta}}||\mathbf{y}-
X\boldsymbol{\bbeta}||_2^2 +
\lambda\sum_{j = 1}^p {\hat w}_j|\bbeta_j|,
\end{equation}
where the weights ${\hat w}_j=(|{\hat \bbeta}_j^0|)^{-\gamma}$
($j=1,\ldots,p$), with $\gamma$ is a positive constant and ${\hat
{\boldsymbol{\bbeta}}}^0$ is an initial consistent estimate of ${\boldsymbol \bbeta}^{*}$. We
recall here that $\hat{\mathbf{\bbeta}}_{\mbox{Lasso}}$ is the solution to a similar equation (\ref{equ2})
in which ${\hat w}_j=1$ for all $j$.

The second most drawback of the Lasso (and also AdaLasso or $\ell_1$
penalization methods) is its poor performance when there are highly
correlated predictors. Under high dimensionality, the situation is
particularly dire. Zou and Hastie (2005) showed that the Lasso
estimates are instable when predictors are highly correlated. They
proposed the Elastic Net (Enet) for variable selection, which
combines $\ell_1$ and $\ell_2$ penalties. El Anbari and Mkhadri
(2008) proposed a procedure called Elastic Corr-Net (Cnet) which combines the
$\ell_1$ and the correlation based penalty of Tutz and Ulbricht
(2009). Daye and Jeng (2009) proposed a slightly similar approach
called the Weighted Fusion (WFusion). These two approaches can
incorporate information redundancy among correlated predictors for
estimation and variable selection.  Numerical studies have shown
that Cnet and WFusion outperform the Lasso and Enet in certain situations. In the same
setting, Hebiri and van De Geer (2010) considered the Smooth-Lasso procedure
(S-Lasso), a modification of the Fused-Lasso procedure (Tibshirani et
al. 1998), in which a second $\ell_1$ Fused penalty is replaced by
the smooth $\ell_2$ norm penalty. The general formulation
englobing all the four latter approaches, called the Generalized
Ridge Lasso (Gril) estimator, can be defined by
\begin{equation}\label{equ3}
{\hat
\bbeta}_{\mbox{Gril}}(\lambda_1,\lambda_2)=\arg\min_{\boldsymbol{\bbeta}}||\mathbf{y}-
X\boldsymbol{\bbeta}||_2^2+\lambda_1||\boldsymbol{\bbeta}||_1
+\lambda_2\boldsymbol{\bbeta}^t{\mathbf Q}\boldsymbol{\bbeta},
\end{equation}
where ${\mathbf Q}$ is a positive semi-definite matrix. A similar
formulation was cited in Daye and Jeng (2009) and Hebiri and  van De Geer (2010) in
regression problem and in Clemmensen et al. (2008) in classification
problem. Moreover, the computation of the estimates of the
parameters of Gril procedure can be obtained efficiently via a
modification of LARS algorithm (Efron et al. 2004).

The Gril estimator (Enet in particular) resolves the collinearity
problem of Lasso, and AdaLasso estimator possesses the oracle property
of SCAD. However, in high dimensional setting, the Gril misses the
oracle property, while AdaLasso estimates are instable because of bias
problem of Lasso. Recently, Zou and Zhang (2009) proposed the
adaptive Elastic Net (AdaEnet) that combines the strengths of
$\ell_2$ norm and the adaptive weighted $\ell_1$ shrinkage. They
established the oracle property of the AdaEnet when the dimension
diverges with the sample size. Independently, Ghosh (2007) proposed
the same AdaEnet, but he specially focused on the grouped selection
property of AdaEnet along with its model selection complexity.

Despite its popularity, Enet (an also AdaEnet) has been critized for being
inadequate, notably in situations in which additional structural knowledge
about predictors should be taken into account (cf. Bondel and Reich 2008,
El anbari and Mkhadri 2008, Daye and Jeng 2009, Hebiri and van De Geer 2010,
Slawski et al. 2010 and She 2010). To this end, these authors complement
$\ell_1-$regularized with a second regularized based on the total variation
or the quadratic penalty. The former aims at the explicit inclusion of structural
knowledge about predictors, while the latter aims at taken into account some
type of correlation between predictors. The experimental results of these alternatives
have shown that Enet performs worse in grouping highly correlated predictors. But,
similar to Enet, these new estimators are asymptotically biased because of the
$\ell_1$ component in the penalty and they cannot achieve selection consistency and
estimation efficiency simultaneously.

Therefore, there is a need to develop methods that take into account of additional
structural information of predictors and have the oracle property. In the same spirit
of AdaEnet, we propose the adaptive Gril (\mbox{AdaGril}) that
penalizes the least square loss using a mixture of weighted $\ell_2$
norm and the adaptive weighted $\ell_1$ penalty. We first highlight
the grouped selection property for AdaCnet method (one type of AdaGril)
in the equal correlation case, meaning that it selects or drops highly correlated
predictors together. Under weak
conditions, as in Zou and Zhang (2009), we study its asymptotic
properties when the dimension diverges with the sample size. In
particular, we show that the AdaGril enjoys the oracle property
with a diverging number of predictors. Moreover, we show that AdaGril
estimator achieves a Sparsity Inequality, i. e., a bound in terms
of the number of non-zero components of the 'true' regression coefficient.
This bound is obtained under a similar weak Restricted Eigenvalue (RE)
condition used for Lasso. Finally, a detailed
experimental performance comparison of different Gril estimators is
considered. 

In Section 2, we focus the Cnet method and sketch briefly other Gril estimators.
 A computational algorithm to approach their solutions  is presented and
we briefly summarize some of their statistical properties obtained in
fixed dimensional setting. In Section 3, we define the Adaptive Gril
estimator and begin by showing the property
of grouping effect of AdaCnet in equal correlation case. Then, we establish
the Statistical asymptotic theory of the AdaGril when the dimension diverges,
including the oracle property. We end by showing that AdaGril achieves a Sparsity
Inequality.
Computational aspects of adaptive Gril is discussed in Section 4.
A detailed simulation study is performed in Section 5,
which illustrates the performance of particular three cases of
Gril and AdaGril estimators in relation to AdaEnet estimator.
A brief discussion is given in Section 6. All technical proofs are provided in Section 7.
\section{Different Gril estimators}
In this section, we present a brief introduction of our alternative to Enet,
called \textit{the Elastic Corrnet (Cnet)}, which takes into account the correlation
between predictors in the quadratic penalty. Two other competitor Gril estimators
are presented and their statistical properties are summarized.
\subsection{Doubly regularized techniques}
Suppose that the predictors are
$\mathbf{x}_i=(x_{i1},\ldots,x_{ip})$ and response values $y_i$, for
$i=1,\ldots,n$. 

Apart from lack of consistency, it
is well known that Lasso has two limitations; for example a) Lasso
does not encourage grouped selection in the presence of high
correlated covariates and b) for $p>n$ case Lasso can select at most
$n$ covariates. To overcome these limitations, Zou and Hastie (2005)
proposed elastic net which combines both ridge $(\ell_2)$ and Lasso
$(\ell_1$) penalties.
So,
Enet procedure corresponds to the Gril estimator with
$\mathbf{Q}={\bf I}_n$, where ${\bf I}_n$ is the $n$x$n$ identity
matrix.

Despite its popularity, Enet (an also AdaEnet) has been critiqued for being
inadequate, notably in situations in which additional structural knowledge
about predictors should be taken into account (cf. Bondel and Reich 2008,
El anbari and Mkhadri 2008, Daye and Jeng 2009, Hebiri and van De Geer 2010,
Slawski et al. 2010 and Shen 2010). To this end, these authors complement
$\ell_1$ regularized with a second regularized based on the total variation
or the quadratic penalty. The former aims at the explicit inclusion of structural
knowledge about predictors, while the latter aims at taken into account some
type of correlation between predictors. One example of the latter, is
the Elastic Corr-Net (Cnet) (EL Anbari and Mkhadri 2008) which is a
modification of Enet in which the ridge penalty term is replaced by
the correlation based penalty term $P_c(\boldsymbol{\bbeta})$ defined by
$$
P_{c}(\boldsymbol{\bbeta)} = \sum_{j=1}^{p-1} \sum_{j > i} \left \{
\frac{(\bbeta_{i} - \bbeta_{j})^{2}}{1 - \rho_{ij}} + \frac{(\bbeta_{i}
+ \bbeta_{j})^{2}}{1 + \rho_{ij}}\right\},
$$
where $\rho_{ij}=\mathbf{x}_i^t\mathbf{x}_j$ denotes the (empirical)
correlation between the $i$th and the $j$th predictors. The
correlation based penalty $P_c(\boldsymbol{\bbeta})$, introduced by Tutz and
Ulbricht (2009), will encourages grouping effect for highly
correlated variables. This penalty can be written in a simple
quadratic form
\begin{eqnarray*}
P_c(\boldsymbol{\bbeta}) = \boldsymbol{\bbeta}^t\mathbf{L}\boldsymbol{\bbeta},
\end{eqnarray*}
where $\mathbf{L} = (\ell_{ij})_{1\leq i,j \leq p}$ is a positive
definite matrix with general term, assuming that $\rho_{ij}^{2}\neq
1$ for $i\neq j$,
\begin{equation}\label{equ5}
    \ell_{ij} =
    \left\{
    \begin{array}{ll}
    2\sum_{s\neq i} \frac{1}{1 - \rho_{is}^{2}},   &
 i = j\\
    -2\frac{\rho_{ij}}{1-\rho_{ij}^{2}},  &
i \neq j.
    \end{array}
    \right.
\end{equation}
Hence, Cnet is a particular Gril estimator with the weighted
matrix $\mathbf{Q}$ defined by (\ref{equ5}). Cnet provided a good performance
in simulations and real applications specially for highly correlated predictors.

We can mention also the Weighted Fusion (WFusion) (Daye and Jeng 2009)
and the Smooth-Lasso (S-Lasso) (Hebiri and van de Geer 2010) as alternatives to
Cnet. In the former, the correlation based penalty is replaced by a modified weighted penalty
${\tilde P}_c(\boldsymbol{\bbeta})= \sum_{j > i}w_{ji}(\bbeta_i-s_{ij}\bbeta_j)^2,$ where
$ w_{ji}=|\rho_{ij}|^{\gamma}/(1-|\rho_{ij}|)$,
$s_{ij}=\mbox{sgn}(\rho_{ij})$ the sign of $\rho_{ij}$ and $\gamma>0$
is a tuning parameter. While, the latter
is a
modification of the Fused-Lasso procedure (Tibshirani et al. 1998),
in which a second $\ell_1$ Fused penalty is replaced by the smooth
$\ell_2$ norm penalty. This quadratic term helps to tackle situations where the
regression vector is structured such that its coefficients vary slowly.
Surprisingly, this simple modification leads to good performance, specially
when the regression vector is 'smooth', i. e., when the variations between successive
coefficients of the unknown parameter of the regression are small.

\subsection{A computational algorithm}
In this section we propose a modification of the Elastic-Net
algorithm for finding a solution of the penalized least squares
problem (\ref{equ3}) of the Gril. The main idea is to transform the
Gril problem into an equivalent Lasso problem on the augmented data
(cf. Zou and Hastie, 2005). Let
\begin{eqnarray}\label{augdata}
\mathbf{\tilde{X}}_{(n + p)\times p} = \left(
                                                             \begin{array}{c}
                                                               \mathbf{X} \\
                                                               \sqrt{\lambda_2}\mathbf{L}^t \\
                                                             \end{array}
                                                           \right), \hspace{5mm} \tilde{\mathbf{y}}_{(n + p)} = \left(
                                                                     \begin{array}{c}
                                                                       \mathbf{y} \\
                                                                       \mathbf{0} \\
                                                                     \end{array}
                                                                   \right) \hspace{3mm}\mbox{and} \hspace{3mm} \mathbf{\tilde{\varepsilon}}_{(n + p)} = \left(
                                                                     \begin{array}{c}
                                                                       \mathbf{\varepsilon} \\
                                                                       \ -\sqrt{\lambda_2}\mathbf{L}^t\bbeta^{\ast} \\
                                                                     \end{array}
                                                                   \right),
\end{eqnarray}

where $\mathbf{Q}$ is a real symmetric semi positive-definite square
matrix with Choleski decomposition $\mathbf{Q} =
\mathbf{L}\mathbf{L}^t$ and $\mathbf{L}=\mathbf{Q}^\frac{1}{2}$.
The Gril estimator is defined as
\begin{eqnarray*}
\hat{\boldsymbol{\bbeta}}=\arg\min_{\boldsymbol{\bbeta}}
\|\tilde{\mathbf{y}} - \mathbf{\tilde{X}}\boldsymbol{\bbeta}\|^{2}_{2} +
\lambda_1\|\boldsymbol{\bbeta}\|_{1}.
\end{eqnarray*}
The latter result is a consequence of
simple algebra, and it motivates the following comment on the Gril method. \\
{\bf Remark 1}. The Gril estimates can be computed via the Lasso
modification of the LARS algorithm. For a fixed $\lambda_2$, it constructs
at each step, which corresponds to a value of $\lambda_1$, an
estimator based on the correlation between covariates and the
current residue. Then for a fixed $\lambda_2$, we obtain the evolution of
the Gril estimator coefficient values when
 $\lambda_1$ varies. It provides the coefficient regularization paths of the
 Gril estimator which are  piecewise linear (Efron et al., 2004).
 Consequently, the Gril algorithm requires the same order of magnitude of computational effort
 as the OLS estimate via the Lasso modification of the LARS algorithm. \\
 {\bf Remark 2}. If $p>n$, it is well known that LARS and its Lasso versions can select
 at most $n$ variables before it puts all coefficients to nonzero.
 Now, applied LARS to augmented data $(\mathbf{\tilde{y}}, \mathbf{\tilde{X}})$,
 the lasso modification of the LARS algorithm is able to select all the $p$
 predictors in all situations. So the first limitation of the Lasso is
 easily surmounted. Moreover, the variable selection is performed in a
 fashion similar to the Lasso.
\subsection{Statistical properties of Different Gril estimators}
The model is assumed to be sparse, i. e. most the regression
coefficients of $\boldsymbol{\bbeta}^{*}$ are  exactly zero corresponding to
predictors that are irrelevant to the response. Without loss of
generality, we assume that the $q$ first components of vector
$\boldsymbol{\bbeta}^{*}$ are non-zero. We briefly summarizes in this section the
classical properties of model selection consistency of particular
Gril estimators.

Yuan and Lin (2007) are the first to give a necessary and sufficient
condition on the  generating covariance matrices for the Elastic net
to select the true model when $q$ and $p$ are fixed. The latter is
called the Elastic Irrepresentable Condition (EIC) which is an
extension of the Irrepresentable Condition (IC), defined in Zhao and
Yu (2006), for Lasso's model selection consistency. For the general
scaling of $q, p$ and $n$,  Jia and Yu (2010) give conditions on
the relationship between $q, p$ and $n$ such that EIC guarantees the
Elastic net's model selection consistency. Moreover, they showed that
EIC is weaker than IC. In the same spirit, consistency properties
and asymptotic normality are established when $p\le n$ for WFusion
(Daye and Jeng 2009). For high dimensional setting $p>n$, Hebiri and van De Geer (2010)
established recently variable selection consistency results for their Quadratic estimator,
which corresponds exactly to our Gril estimator. They showed that Gril estimator
achieves a Sparsity Inequality, i. e., a bound in terms of the number non-zero
components of the 'true' vector regression. The latter result for $n>p$ is extended to
AdaGril estimator in the next Section and its oracle properties are detailled
when $p$ diverges.
\section{The adaptive Gril estimator}
Now a revised version of Gril estimator, called AdaGril, is proposed
by incorporating the adaptive weights in the $\ell_1$ penalty of
equation (\ref{equ3}). So, AdaGril is a combination of Gril and
AdaLasso. We first assume that ${\hat{\boldsymbol{\bbeta}}}^{0}$ is an initial
estimator of $\boldsymbol{\bbeta}^{*}$ which is a root $n$-consistent. For example, we
can choose ${\hat{\boldsymbol{\bbeta}}}_{\mbox{ols}}$ or
 ${\hat{\boldsymbol{\bbeta}}}_{\mbox{Gril}}$, and we construct the weights by
\begin{equation} \label{equ8}
\hat{\omega}_{j}=(|\hat{\bbeta}_{j}(\mbox{Gril})|)^{-\gamma},
\hspace*{3mm} j = 1, ..., p,
\end{equation}
where $\gamma$ is a positive constant. \textcolor{black} {Let $(q_{1}, ..., q_{p})$ be the diagonal elements of $\mathbf{Q}$ and let the $p\times p$ matrix $\mathbf{N}$ defined by
$$
 \mathbf{N} = \mbox{diag}\left(1 + \frac{\lambda_{2}q_{1}}{n},1 + \frac{\lambda_{2}q_{2}}{n},
\ldots,1 + \frac{\lambda_{2}q_{p}}{n}\right).
$$}
Then, the adaptive Gril
estimates are defined by
\begin{equation} \label{equ9}
\hat{\bbeta}(\mbox{AdaGril}) = \textcolor{black} {\mathbf{N}} \left\{\arg\min_{\boldsymbol{\bbeta}} \|\mathbf{y} -
\mathbf{X}\boldsymbol{\bbeta}\|_{2}^{2} + \lambda_2\boldsymbol{\bbeta}^t\mathbf{Q}
\boldsymbol{\bbeta} +
\lambda_1^{*}\sum_{j=1}^{p} \hat{\omega}_{j}|\bbeta_{j}|\right\}.
\end{equation}
To overcome dividing by zeros, we can choose
$\hat{\omega}_j=(|\hat{\bbeta}_{j}(\mbox{Gril})+1/n|)^{-\gamma}$ or
$\hat{\omega}_j=\infty$.

Now, it is clear that AdaGril combines the strengths of Ridge
regression and AdaLasso. So, AdaGril will avoid both the problem of
collinearity and bias problem of Lasso in high dimensional setting.
The tuning parameters $\lambda_1^{*}$ and $\lambda_1$ are directly
responsable of sparsity of the estimates and are allowed to be
different. While the same value of $\lambda_2$ is used for Gril and
AdaGril estimators, because the quadratic norm in the $\ell_2$
penalty leads to the same kind of contribution in both estimators.

\subsection{The grouping effect of AdaCnet}
Grouping effect is expressed when the regression coefficients of a group of highly
correlated variables tend to be equal (up to a change of sign if
negatively correlated). Similar to Cnet estimator, the AdaCnet estimator has
the natural tendency of grouping each pair of regression coefficients according
to their correlations. We establish in the following lemma the
   grouping effect of AdaCnet in the case of equal correlations.
\begin{lem}\label{mylemma1}
Given data $(\mathbf{y},\mathbf{X})$, where $\mathbf{X}=
(\mathbf{x}_{1}|...|\mathbf{x}_{p})$ and parameters $(\lambda_{1}^{*},
\lambda_{2})$, the response is centered and the predictors
$\mathbf{X}$ standardized. Let $\hat{\mathbf{\bbeta}}(\lambda_{1}^{*},
\lambda_{2})$ be the AdaCnet estimate.

If
$\hat{\bbeta}_{i}(\lambda_{1}^{*},
\lambda_{2})\hat{\bbeta}_{j}(\lambda_{1}^{*}, \lambda_{2}) > 0$ and
$\rho_{kl} = \rho, \hspace*{3mm}\mbox{for all} \hspace*{1mm} (k, l)$
, then
\begin{eqnarray*}\label{eq5}
\frac{1}{\|\mathbf{y}\|_{2}}\left|\hat{\bbeta}_{j} -
\hat{\bbeta}_{i}\right| & \leq & \frac{1 - \rho^{2}}{2(p + \rho - 1)\lambda_{2}}
\left[\sqrt{2(1 - \rho)}
+\frac{\gamma\lambda_{1}^{*}}{\|\mathbf{y}\|_2\min(|\hat{\bbeta}_i^{�}|,|\hat{\bbeta}_j^{�}|)^{\gamma+1}}|\hat{\bbeta}_i^{�}-\hat{\bbeta}_j^{�}|\right]
\end{eqnarray*}
\end{lem}
{\bf Remark 3.} We note that $\gamma=0$ leads the grouping effect of the Cnet as a special case.
We also observe that the grouping effect has contributions not only from quadratic type penalty
but also from $L_1$ type adaptive penalty. However if $\lambda \rightarrow 0$, then it is not
possible to capture any grouping effect from only the $L_1$ type adaptive penalty. Moreover, when
considering $\hat{\bbeta}_j^{�}$ as univariate OLS estimates with
$\min(|\hat{\bbeta}_i^{�}|,|\hat{\bbeta}_j^{�}|)\ge 1$, the latter becomes
$$
\frac{1}{\|\mathbf{y}\|_{2}}\left|\hat{\bbeta}_{j} -
\hat{\bbeta}_{i}\right|  \leq  \frac{1 - \rho^{2}}{2(p + \rho - 1)\lambda_{2}}
\left(2+\gamma\lambda_{1}^{*} \right)\sqrt{2(1 - \rho)}.
$$
\subsection{Model selection consistency for AdaGril when $p$ diverges}
The oracle properties of the adaptive Elastic Net is provided in Ghosh (2007)
for $p\le n$. But, a detailed and much more elaborate discussion of
the oracle properties of the adaptive elastic net is provided in Zou and
Zhang (2009). In this section and as in Zou and Zhang (2009) we
establish the oracle properties of the AdaGril estimator when $p$ diverges
(i. e. $p(n)=n^{\nu}, 0\le\nu<1$). Moreover, we provide a bound on the mean  squared sparsity
inequality, that is a bound on the mean squared risk that takes into
account the sparsity of the oracle regression vector $\boldsymbol{\bbeta}$.

\subsubsection{Mean Sparsity Inequality}
Now we establish the mean sparsity inequality achieved by the
AdaGril estimator. For this purpose, we need the following
assumption on the minimum and the maximum eigenvalues of the
semi-positive definite matrices $\mathbf{X}^t\mathbf{X}$ and $\mathbf{Q}$, respectively.\\
(C1) Let $\lambda_{\min}(\mathbf{M})$ and $\lambda_{max}(\mathbf{M})$
denote the minimum and the maximum eigenvalues of a semi-positive
definite matrix $\mathbf{M}$, respectively. Then we assume
$$b\leq
\lambda_{\min}(\frac{1}{n}\mathbf{X}^t\mathbf{X})\leq
\lambda_{max}(\frac{1}{n}\mathbf{X}^t\mathbf{X})\leq B
$$
and
$$
d\leq \lambda_{\min}(\mathbf{Q})\leq \lambda_{max}(\mathbf{Q})\leq D
$$
where $b, B, d$ and $D$ are constants so that $b, B>0$ and $d, D
\geq 0$.

Now, given the data $(\mathbf{y}, \mathbf{X})$, let
$\hat{\boldsymbol{\omega}} = (\hat{\omega}_{1}, ..., \hat{\omega}_{p})$
be a vector whose components are all non-negative and can depend on
$(\mathbf{y}, \mathbf{X})$. Define
$$
\hat{\boldsymbol{\bbeta}}_{\hat{\boldsymbol{\omega}}}(\lambda_2, \lambda_{1}^{*}) =
\left\{\arg\min_{\boldsymbol{\bbeta}} \|\mathbf{y} - \mathbf{X}\boldsymbol{\bbeta}\|_{2}^{2} +
\lambda_2 \boldsymbol{\bbeta}^t\mathbf{Q}\bbeta + \lambda_{1}^{*}\sum_{j=1}^{p}
\hat{\omega}_{j}|\bbeta_{j}|\right\}
$$
for non-negative parameters $\lambda_{1}^{*}$ and $\lambda_2$. If
$\hat{\omega}_{j} = 1$ for all $j$, we denote
$\hat{\boldsymbol{\bbeta}}_{\hat{\boldsymbol{\omega}}}(\lambda_2,\lambda_{1}^{*})$ by
$\hat{\boldsymbol{\bbeta}}(\lambda_2, \lambda_{1}^{*})$ for convenience. The assumption (C1)
assume a reasonably good behavior of both the predictor and the
weight matrices (cf Portnoy 1984).
\begin{thm}\label{mytheo1}
If we assume the model (\ref{equ1}) and Assumption (C1), then
$$
I\!\!E\left(\|\hat{\bbeta}_{\hat{\mathbf{\omega}}}\left(\lambda_2,
\lambda_{1}^{*}\right) - \bbeta^{*}\|_{2}^{2}\right) \leq
4\frac{\lambda_2^{2}D^{2}\|\bbeta^{*}\|_{2}^{2} + Bpn\sigma^{2} +
\lambda_{1}^{*2}I\!\!E\left(\sum_{j=1}^{p}
\hat{\omega}_{j}^{2}\right)}{\left(bn + \lambda_2 d\right)^{2}}
$$
In particular, when $\hat{\omega}_{j} = 1$ for all $j$ and if we note $\lambda_{1}^{*}$ by $\lambda_{1}$, we obtain
the mean sparsity inequality for the Gril estimator.
$$
\!\!E\left(\|\hat{\boldsymbol{\bbeta}}\left(\lambda_{2}, \lambda_{1}\right) -
\boldsymbol{\bbeta}^{*}\|_{2}^{2}\right) \leq
4\frac{\lambda_2^{2}D^{2}\|\boldsymbol{\bbeta}^{*}\|_{2}^{2} + Bpn\sigma^{2} +
\lambda_{1}^{2}p}{(bn + \lambda_2 d)^{2}}
$$
\end{thm}
The latter risk bounds in Theorem \ref{mytheo1} are non-asymptotic. It implies
that, under assumptions (C1)-(C6) defined below,
$\hat{\boldsymbol{\bbeta}}(\lambda_{1}^{*},\lambda_2)$ is a root-($n/p$)-consistent estimator
(cf. Fan and Peng (2004) for SCAD and Zou and Zhang (2009 ) for
AdaEnet). So, the construction of the adaptive weight by using the
Gril is appropriate.
\subsubsection{Oracle properties}
To establish the oracle properties, we need the same following
assumptions used in
Zou and Zhang (2009).\\
(C2) $\lim_{n\rightarrow\infty} \frac{\max_{i=1, 2, ..., n}
\sum_{j=1}^{p} x_{ij}^{2}}{n} = 0.$\\
(C3) $I\!\!E[|\varepsilon|^{2 + \delta}] <\infty$ for some $\delta
>0$.\\
(C4) $\lim_{n\rightarrow\infty} \frac{\log(p)}{\log(n)} = \nu$ for
some $0\leq \nu < 1$.\\\\
To construct the adaptive weights ($\hat{\omega}$), we take a fixed
$\gamma > \frac{2\nu}{1 - \nu}$. In our numerical studies we let
$\gamma = [\frac{2\nu}{1 - \nu}] + 1$ to avoid tuning on $\gamma$ as in Zou and Zhang (2009).
Once $\gamma$ is chosen, we choose the regularization parameters
according to the following conditions\\\\
(C5) $$\textcolor{black}{\lim_{n\rightarrow\infty} \frac{\lambda_2q_{i}}{n} = 0, \hspace*{3mm} \mbox{for all}\hspace*{3mm} i = 1, ..., p}, \hspace*{3mm}
\lim_{n\rightarrow\infty} \frac{\lambda_1}{\sqrt{n}} = 0,$$
\hspace*{1cm} and
$$\lim_{n\rightarrow\infty} \frac{\lambda_1^{*}}{\sqrt{n}} = 0,\hspace*{3mm}
\lim_{n\rightarrow\infty} \frac{\lambda_1^{*}}{\sqrt{n}}n^{\frac{(1 -
\nu)(1 + \gamma) - 1}{2}} = \infty.$$
\\\\
(C6) $$\lim_{n\rightarrow\infty}
\frac{\lambda_2}{\sqrt{n}}\sqrt{\sum_{j\in \mathcal{A}} \bbeta_{j}^{*2}} =
0,\hspace*{3mm} \lim_{n\rightarrow\infty}\left(\frac{\lambda_{1}^{*}}{n}\right)^{\frac{2\gamma}{1 + \gamma}}\frac{p\|\bbeta^{*}\|_{2}^{2}}{\lambda_{1}^{*2}} = 0,\hspace*{3mm}\lim_{n\rightarrow\infty}
\min(\frac{n}{\lambda_{1}\sqrt{p}},
\left(\frac{\sqrt{n}}{\sqrt{p}\lambda_1^{*}}\right)^{\frac{1}{\gamma}})(\min
_{j\in \mathcal{A}} |\bbeta_{j}^{*}|)\rightarrow\infty.$$
\begin{thm}\label{mytheo2}
Let us write $\boldsymbol{\bbeta}^{*} = (\boldsymbol{\bbeta}_{\mathcal{A}}^{*}, 0)$
and define
\begin{equation} \label{equ10}
\tilde{\boldsymbol{\bbeta}}_{\mathcal{A}}^{*} = \arg\min_{\boldsymbol{\bbeta}}
\left\{\|\mathbf{y} - \mathbf{X}_{\mathcal{A}}\boldsymbol{\bbeta}\|_{2}^{2} +
\lambda_2\boldsymbol{\bbeta}^t\mathbf{Q}_{\mathcal{A}}\boldsymbol{\bbeta} +
\lambda_1^{*}\sum_{j\in\mathcal{A}}
\hat{\boldsymbol{\omega}}_{j}|\bbeta_{j}|\right\}.
\end{equation}
Then, under the assumptions (C1)-(C6) and with probability tending
to $1$, \textcolor{black}{$(\mathbf{N}_{\mathcal{A}}\tilde{\boldsymbol{\bbeta}}_{\mathcal{A}}^{*}, 0)$} is solution to
(\ref{equ9}).
\end{thm}
Theorem \ref{mytheo2} provides an asymptotic characterization of the solution
to the adaptive Gril criterion. It demonstrates that the Adaptive
Gril estimator is as efficient as an oracle one. Moreover, it is
helpful in the proof of Theorem 3.3 below.
\begin{thm}\label{mytheo3}
Under conditions (C1)-(C6), the adaptive Generalized Ridge Lasso has
the oracle property, that is, the estimator
$\hat{\boldsymbol{\bbeta}}(\mbox{AdaGril})$ must satisfy:
\begin{enumerate}
  \item Consistency in selection : $\mbox{Pr}\left(\{j : \hat{\boldsymbol{\bbeta}}(\mbox{AdaGril})_{j} \neq 0 \} = \mathcal{A}\right)\rightarrow 1,$
  \item \textcolor{black}{Asymptotic normality : $\alpha^t\Sigma_{\mathcal{A}}^{\frac{1}{2}}\left(I +
\lambda_2\Sigma_{\mathcal{A}}^{-1}\mathbf{Q}_{\mathcal{A}}\right)\mathbf{N}_{\mathcal{A}}^{-1}\left(\hat{\bbeta}(\mbox{AdaGril})_{\mathcal{A}}-
\boldsymbol{\bbeta}_{\mathcal{A}}^{*}\right)
  \rightarrow_{d}\mbox{N}(0, \sigma^{2})$, where $\mathbf{X}_{\mathcal{A}}$, $\mathbf{Q}_{\mathcal{A}}$ and $\mathbf{N}_{\mathcal{A}}$ are sub-matrices obtained by extracting the columns of $\mathbf{X}$, $\mathbf{Q}$ and $\mathbf{N}$ respectively according to the indices in $\mathcal{A}$, $\Sigma_{\mathcal{A}} = \mathbf{X}_{\mathcal{A}}^t\mathbf{X}_{\mathcal{A}}$
  and $\mathbf{\boldsymbol{\alpha}}$ is a vector of norm $1$.}
\end{enumerate}
\end{thm}
Theorem \ref{mytheo3} provides the selection consistency and asymptotic
normality of AdaGril when the number of parameters diverges. So,
AdaGril estimator enjoys the oracle property of SCAD in high
dimensional setting. As a first special case and taking $\mathbf{Q} = \mathbf{I}$, we obtain the asymptotic normality of the Adaptive elastic net:

$$\alpha^t\frac{I +
\lambda_2\Sigma_{\mathcal{A}}^{-1}}{1 + \frac{\lambda_{2}}{n}}\Sigma_{\mathcal{A}}^{\frac{1}{2}}\left(\hat{\bbeta}(\mbox{AdaEnet})_{\mathcal{A}}-
\boldsymbol{\bbeta}_{\mathcal{A}}^{*}\right)
  \rightarrow_{d}\mbox{N}(0, \sigma^{2}).$$

  Taking $\lambda_{2} = 0$, we obtain the asymptotic normality of the Adaptive Lasso as a second special case:

  $$\alpha^t\Sigma_{\mathcal{A}}^{\frac{1}{2}}\left(\hat{\bbeta}(\mbox{AdaLasso})_{\mathcal{A}}-
\boldsymbol{\bbeta}_{\mathcal{A}}^{*}\right)
  \rightarrow_{d}\mbox{N}(0, \sigma^{2}).$$
\subsection{Sparsity inequality for AdaGril estimator}
Now we establish a sparsity inequality (SI) achieved by $\hat{\bbeta}(\mbox{AdaGril})$, that is a bound on $L_2$ and $L_1$ error estimation, in terms of the number of non-zero components of the 'true' coefficient vector $\boldsymbol{\bbeta}^{\ast}$. Here the second parameter $\lambda_2$ is not free, but it depends on the parameter $\lambda_1^{*}$ which is fixed as a function of $(n,p,\sigma)$. Moreover, the Gril estimator (instead of OLS or Lasso estimator) is used as the initial estimator for the adaptive Gril method. Finally, our result of sparsity inequalities are obtained under a similar assumption on the Gram matrix used by the Lasso (cf. Bickel et al. 2009).
Let us now establish the assumptions needed. \\
\textbf{Assumption} RE.
\textit{There is a constant $\psi > 0$ such that, for any $\mathbf{z} \in \mathbb{R}^{p}$  that satisfies $\sum_{j \in \mathcal{A}^{c}} |\mathbf{z}_{j}| \leq 4\max\left\{\frac{2}{\eta}, 1\right\}\sum_{j \in \mathcal{A}} |\mathbf{z}_{j}|$, we have}
$$\mathbf{z}^t\mathbf{K}\mathbf{z} \geq \psi \sum_{j \in \mathcal{A}}\mathbf{z}_{j}^{2},$$

\textit{where $\mathbf{K} = \mathbf{\tilde{X}}^t\mathbf{\tilde{X}}$ and $\eta = \min_{j \in \mathcal{A}}(|\boldsymbol{\bbeta}_{j}^{*}|).$}\\
Let $\hat{\bbeta}(\mbox{Gril})$ and $\hat{\bbeta}(\mbox{AdaGril})$ denote the Gril estimator and the adaptive Gril estimator, respectively. Here, the weights of the adaptive estimator are estimated from the Gril estimator $\hat{\bbeta}(\mbox{Gril})$:
 \begin{equation} \label{poids}
\hat{\omega}_{j} = \max(\frac{1}{|\hat{\bbeta}(\mbox{Gril})_{j}|}, 1)\hspace*{3mm}\mbox{for all}\hspace*{3mm} j = 1, ..., p.
\end{equation}
We note that Zhou et al. (2009) have also considered the same weights in their analysis of the adaptive Lasso for high dimensional regression and Gaussian graphical models. These weights are easy to manipulate in our proof of the next Theorem than the classical weights (\ref{equ8}).
\begin{thm}\label{oracle.theo}
Given data $(\mathbf{y}, \mathbf{X})$. Let $s = |\mathcal{A}|, \eta = \min_{j \in \mathcal{A}}(|\boldsymbol{\bbeta}_{j}^{*}|)$ and $\varphi \in (0,1)$. We define our tuning parameter $\lambda_1^{*}$ and $\lambda_2$ as follows
\begin{equation*} \label{oraclebis3.theo}
\lambda_1^{*} = 8\sqrt{2}\sigma\sqrt{\frac{\log(p/\varphi)}{n}} \quad \text{and} \quad \lambda_{2} = \frac{\lambda_1^{*}}{8\|\mathbf{Q}\boldsymbol{\boldsymbol{\bbeta}^{\ast}}\|_{\infty}}.
\end{equation*}
Now, let $\delta = 8\psi^{-1}\lambda_1^{*}s.$ and for $0 <\theta \leq 1$, consider the set $\Gamma = \{ \max_{j=1, ..., p} 2|U_{j}| \leq \theta\lambda_{1}\}$ with $U_{j} = n^{-1} \sum_{i=1}^{n} x_{i,j}\varepsilon_{i}$.
Therefore, if assumption RE holds and in addition $\eta \geq 2\delta$, then with probability greater than $1 - \varphi$ on the $\Gamma$ set, we have
\begin{eqnarray*}\label{oraclebis5.theo}
\|\mathbf{X}\boldsymbol{\boldsymbol{\bbeta}^{\ast}} - \mathbf{X}\hat{\bbeta}(\mbox{AdaGril})\|_{2}^{2}
& \leq & 4\psi^{-1}\lambda_1^{*2}\left(\max\left\{\frac{2}{\eta}, 1\right\}\right)^{2} s,
\end{eqnarray*}
\begin{eqnarray*}\label{oraclebis6.theo}
\|\boldsymbol{\boldsymbol{\bbeta}^{\ast}} - \hat{\bbeta}(\mbox{AdaGril})\|_{1}
& \leq & 8\psi^{-1}\lambda_1^{*}\left(\max\left\{\frac{2}{\eta}, 1\right\}\right)^{2} s.
\end{eqnarray*}
\end{thm}
The Restricted Eigenvalue (RE) Assumption is widely used in the literature about the variable selection consistency of $\ell_1$-penalized regression methods in high dimension ($p>>n$, see for instance Bickel et al. 2009, Zhou et al. 2009 and Hebiri and van De Geer 2010). On the one hand,
the main difference of our RE assumption with that in Bickel et al. (2009) are in the matrix
$\tilde{K}=\mathbf{X}^t\mathbf{X}+\lambda_2\mathbf{Q}$ and the specified constant $4\max\left\{\frac{2}{\eta}, 1\right\}$ instead of $K_{n}=n^{-1}\mathbf{X}^t \mathbf{X}$ and an arbitrary constant $cte$, respectively. On the other hand, there is a minor difference with the assumption $B(\Theta)$ used in Hebiri and van De Geer (2010). Indeed, the latter authors only need to consider the vectors $\mathbf{z}$ such that ${\sum_{j \notin \Theta}} |\mathbf{z}_{j}| \leq \rho_n\sqrt{{\sum_{j \in \Theta}}\mathbf{z}_j^2}$, where $\rho_n$ is a scalar which depend of $(s,\lambda_1^{*},\lambda_2,\boldsymbol{\bbeta}^{*})$. Moreover, our choice of regularized parameters $(\lambda_1^{*},\lambda_2)$ are relatively similar to that used by Hebiri and van De Geer (2010) in Corollary $1$ for the sparsity inequality of Gril estimator (called in that paper Quadratic estimator). We then refer the reader to the latter reference for more discussions about that choice.
\section{Computation and tuning parameters selection}
In this section we propose a modification of the Gril
algorithm for finding a solution of the penalized least squares
problem (\ref{equ9}) of the AdaGril. The main idea is to transform the
AdaGril problem into an equivalent Gril problem on the augmented data
(cf. Zou and Hastie, 2005). The main steps of the AdaGril algorithm are as follows:

  $1.$ Input: Matrix $\mathbf X$ and $\hat{\boldsymbol{\omega}}$.

  $2.$ Put $\mathbf{x}_j^{**}=\mathbf{x}_j\hat{\omega}_j$ for $j=1,\ldots,p$

  $3.$ Use the Gril algorithm described in the Section 2.3 for computing the AdaGril estimator
  $\hat{\boldsymbol{\bbeta}}(\mbox{AdaGril})$.

In practice, it is important to select appropriate tuning parameters
$(\lambda_1,\lambda_2,\gamma)$ in order to obtain a good prediction
precision. Choosing the tuning parameters can be done via minimizing
an estimate of the out-of-sample prediction error. If a validation
set is available, this can be estimated directly. Lacking a
validation set one can use ten-fold cross validation. Note that
we take a fixed $\gamma = [\frac{2\nu}{1 - \nu}] + 1$ to avoid tuning on $\gamma$. So
there are two tuning parameters in the AdaGril, so we need
to cross-validate on a two dimensional surface. Typically we first
pick a (relatively small) grid values for $\lambda_2$, say $(0,
0.01, 0.1, 1, 10, 100)$. Then, for each
$\lambda_2$, LARS algorithm produces the entire solution path of
the AdaGril. The other tuning parameter is selected by
tenfold CV. The chosen $\lambda_2$ is the one giving the smallest
CV error or generalized cross-validation (GCV). However, Wang, Li and Tsai ($2007$)
showed that for the SCAD method (cf. Fan and Li, $2001$), BIC criterion
is a better tuning parameter selector than GCV and AIC. In our implementations the parameter $\gamma$ is fixed
  for the three adaptive methods (AdaEnet, AdaLasso and AdaGril), while the couple of parameters
$(\lambda_1,\lambda_2)$ is selected using BIC criterion.

\section{Numerical study}
In this section we consider some simulation experiments to evaluate the finite sample performance of
different AdaGril estimators. AdapCnet, AdapWfusion and AdapSlasso methods correspond to adaptive Cnet,
adaptive WFusion and adaptive Smooth-Lasso methods, respectively. These three adaptive versions of AdaGril
are compared with Lasso, AdaLasso and AdaEnet. We consider the first simulated example used in Zou and Zhang (2009). In this example we generate data from the model,
$$y = \mathbf{x}^t\bbeta^{*} + \epsilon,$$

 where $\bbeta^{*}$ is a vector of length $p$ and $\epsilon \sim \mbox{N}(0, \sigma^{2})$, $\sigma \in \{3, 6, 9\}$ and $\mathbf{x} \sim \mbox{N}_{p}(\mathbf{0}, \mathbf{R})$, $\mathbf{R}$ is the correlation matrix whose $(i,j)$th element is $\mathbf{R}_{i,j} = \rho^{|i - j|}$. Results are given for $\rho = 0.5$ and $\rho = 0.75$. This example presents a situation in which the number of parameters depends on the sample size $n$ as follows: $p = p_{n} = [4n^{1/2}] - 5$ for $n= 100, 200, 1000$. The true parameter is
 $$\boldsymbol{\bbeta} = (1, 2, ..., q -1,  q, \underbrace{0,
..., 0}_{p - 3q}, \underbrace{3, ..., 3}_{q}, -1, -2, ..., -q + 1, -q)^t,$$
 where $q = [p_{n}/9]$. For this choice of $n$ and $p$, we have $\nu = \frac{1}{2}$, so we used $\gamma = 3$ for calculating the adaptive weights for all adaptive methods.

 \begin{Large}
\begin{center}
$\ast \ast \ast$ Table $1$ GOES HERE $\ast \ast \ast$
\end{center}
\end{Large}

\begin{Large}
\begin{center}
$\ast \ast \ast$ Table $2$ GOES HERE $\ast \ast \ast$
\end{center}
\end{Large}

\begin{Large}
\begin{center}
$\ast \ast \ast$ Table $3$ GOES HERE $\ast \ast \ast$
\end{center}
\end{Large}

 Table $1$, Table $2$ and Table $3$ summarize the performance of different adaptive and non adaptive methods in terms
 of prediction accuracy, estimation error and variable selection, respectively. Several observations can be made from these tables.\\
$1.$ The adaptive methods outperform the non adaptive ones in terms of prediction and estimation
accuracies, except in two small sample setting cases (i.e. $n=100$ and $200$ for $\sigma=9$).\\
$2.$ In small sample settings, AdaSlasso is the winner in term of prediction accuracy followed by
AdaCnet or Adalasso (except in [$n=100, \sigma=9,\rho=0.5$] where Enet is the winner). However, for large sample,
AdaCnet is the best in term of prediction accuracy followed by Adalasso (except in one case of $[\sigma=9,\rho=0.75]$). \\
$3.$ The AdaSlasso (or Slasso for $n=100$ and $\sigma=6-9$, Table $2$) seems to dominate its competitors in term
of prediction error (i.e. MSE$_{\bbeta}$) in small sample settings ($n=100-200$). It is followed by AdaCnet or Cnet.
However, AdaCnet is by far better than all other method in large sample size $n=1000$ (except the case $\sigma=9$ and $\rho=0.75).$\\
$4.$ When increasing the noise level $\sigma$, the methods behave in the same way by increasing substantially their prediction and error accuracies, and regardless of the sample size $n$ and the correlation coefficient $\rho$.\\
$5.$ From Table $3$, it can be seen that the performance in term of correct selection of all methods increase largely when the sample size increases, and whatever the value of noise level. While, their performance in term
of incorrect selection increase slightly when the noise level increases, and especially in small sample settings.
Moreover, the performance of the adaptive methods, in small settings, is relatively similar and is slightly
better than those of the non adaptive ones (the difference between theme is about $3-5$ percent), and whatever
the values of $\sigma$ and $\rho$. However, in large sample setting, all methods behave in the same way by increasing largely their performance of correct selection of the relevant variables with a little advantage ($3-5$ percent) to AdaCnet and Cnet.

Finally, we can conclude that in this example, the adaptive methods perform better than the non-adaptive ones in terms of variable selection and prediction accuracies, and whatever
the values of $n, \sigma$ and $\rho$. Moreover, the AdaCnet and AdaSlasso outperform largely AdaEnet in quasi different situations.
\textcolor{black}{We have also considered a second example (Example 1 in Zou and Zhang 2009, results not reported here) where the structure of the parameter vector is smooth with small difference between successive coefficients. The
results steal relatively similar to those obtained in example $1$, but with some advantage to AdaSlasso in prediction
accuracy and Slasso in prediction error.}

\textcolor{black}{So, when the structure of the parameter vector is smooth, Slasso and AdaSlasso will have a clear advantage
than its competitors. When this structure is not smooth  and the coefficients have different signs, then
Cnet and AdaCnet seem to work well in this setting. On the other hand, Enet and AdaEnet will give good results in
extreme correlation case ($\rho\approx 1$), while its competitors (Cnet and Wfusion) give good results
when the correlation is moderate. When the correlation is small, Lasso or AdaLasso can do better.}
\section{Discussion}
In this paper we propose AdaGril for variable selection with a diverging number of parameters in the presence of highly correlated variables. AdaGril is a generalization of AdaEnet by replacing the identity
matrix in the $L_{2}$ norm penalty by any positive semi-definite matrix $Q$. Many possible choices of $Q$ are
in the literature. We show that under some conditions on the eigenvalues of $Q$ we can extend results on variable selection consistency and asymptotic normality of the AdaEnet to the AdaGril. Moreover, we show that AdaGril estimator achieves a Sparsity Inequality, i. e.,
a bound in terms of the number of non-zero components of the 'true' regression coefficient. This
bound is obtained under a similar weak Restricted Eigenvalue (RE) condition used for Lasso.
Simulations studies show that some particular cases of AdaGril outperform its competitors. Simulated examples suggests that AdaGril methods
improve both the AdaLasso and AdaEnet. The extension of the AdaGril to generalized linear models (McCullagh and Nelder, $1989$) will be subject to future work.

\section{Proofs}
\begin{proof}
PROOF OF LEMMA \ref{mylemma1}: Let $\hat{\bbeta}= \hat{\bbeta}(\lambda_{1}^{*}, \lambda_{2}) =
\arg\min_{\bbeta}\{L(\lambda_{1}^{*}, \lambda_{2}, \mathbf{\bbeta})\}$, where $$ L(\lambda_{1}^{*}, \lambda_{2}, \mathbf{\bbeta}) = \textcolor{black}{\mathbf{N}}\left\{\|\mathbf{y} -
\mathbf{X}\boldsymbol{\bbeta}\|_{2}^{2} + \lambda_2\boldsymbol{\bbeta}^t\mathbf{Q}
\boldsymbol{\bbeta} +
\lambda_{1}^{*}\sum_{j=1}^{p} \hat{\omega}_{j}|\bbeta_{j}|\right\},$$ and $\mathbf{Q}$ is defined by (\ref{equ5}).
If $\hat{\bbeta}_{i}\hat{\bbeta}_{j} > 0$, then
both $\hat{\bbeta}_{i}$ and
$\hat{\bbeta}_{j}$ are non-zero, and we
have $\mbox{sign}(\hat{\bbeta}_{i}) =
\mbox{sign}(\hat{\bbeta}_{j})$.
 Then $
\hat{\bbeta}$ must satisfies
\begin{equation} \label{eq6}
\frac{\partial L(\lambda_{1}^{*}, \lambda_{2}, \mathbf{\bbeta})}{\partial
\mathbf{\bbeta}}|_{\mathbf{\bbeta} = \hat{\bbeta}} = \mathbf{0}.
\end{equation}
Hence we have
\begin{equation} \label{eq7}
-2\mathbf{x}_{i}^t\{\mathbf{y} -
\mathbf{X}\hat{\mathbf{\bbeta}}\} +
\lambda_{1}^{*}\hat{w}_i\mbox{sign}\{\hat{\bbeta}_{i}\}
+ 2\lambda_{2}\sum_{k=1}^{p} q_{ik}\hat{\bbeta}_{k}=0,
\end{equation}
and
\begin{equation} \label{eqq8}
-2\mathbf{x}_{j}^t\{\mathbf{y} -
\mathbf{X}\hat{\mathbf{\bbeta}}\} +
\lambda_{1}^{*}\hat{w}_j\mbox{sign}\{\hat{\bbeta}_{j}\}
+ 2\lambda_{2}\sum_{k=1}^{p} q_{jk}\hat{\bbeta}_{k}=0,
\end{equation}
Subtracting equation (\ref{eq7}) from (\ref{eqq8}) gives
\begin{eqnarray*}
2(\mathbf{x}_{j}^t - \mathbf{x}_{i}^t)\{\mathbf{y} -
\mathbf{X}\hat{\mathbf{\bbeta}}\} -\lambda_{1}^{*}(\hat{w}_j-\hat{w}_i)\mbox{sign}\{\hat{\bbeta}_{j}\}-
2\lambda_{2}\sum_{k=1}^{p} (q_{jk} -
q_{ik})\hat{\bbeta}_{k}= 0,
\end{eqnarray*}
which is equivalent to
\begin{equation} \label{eq9}
\lambda_2\sum_{k=1}^{p} (q_{jk}
-q_{ik})\hat{\bbeta}_{k} = (\mathbf{x}_{i}^t -\mathbf{x}_{j}^t)\hat{\mathbf{r}}-
\frac{\lambda_{1}^{*}}{2}(\hat{w}_j-\hat{w}_i)\mbox{sign}\{\hat{\bbeta}_{j}\}
\end{equation}
where $\hat{\mathbf{r}}=\mathbf{y} -
\mathbf{X}\hat{\mathbf{\bbeta}}$ is the
residual vector. Since $\mathbf{X}$ is standardized, then
$\|\mathbf{x}_{i} - \mathbf{x}_{j}\|_{2}^{2}=2(1 - \rho_{ij}).$
Because $\hat{\bbeta}$ is the minimizer we
must have
$L\{\lambda_{1}^{*}, \lambda_{2},
\hat{\bbeta}(\lambda_{1}^{*}, \lambda_{2})\} \leq L\{\lambda_{1}^{*},
\lambda_{2}, \mathbf{\bbeta}=\mathbf{0}\},
$ i.e.
\begin{eqnarray}\label{eq10}
\|\hat{\mathbf{r}}\|^{2} +\lambda_{2}\hat{\bbeta}^t\mathbf{Q}\hat{\bbeta} +
\lambda_{1}^{*}\sum_{k=1}^p\hat{w}_k|\hat{\bbeta}_{k}| \leq
\|\mathbf{y}\|_{2}^{2}.
\end{eqnarray}
So $\|\hat{\mathbf{r}}(\lambda_{1}^{*}, \lambda_{2})\|_{2} \leq
\|\mathbf{y}\|_{2}$. \\
Now, we apply the mean value Theorem, as in Ghosh (2007, Proof of Theorem 3.3), to the function $g(x)=x^{-\gamma}$, we have $|g(x)-g(y)|= |g'(c)||x-y|$ for some $c\in[\min(x,y),\max(x,y)]$. Hence we obtain
\begin{eqnarray*}
|\hat{w}_i-\hat{w}_j|&=&| |\hat{\bbeta}_i^{�}|^{-\gamma}-|\hat{\bbeta}_j^{�}|^{-\gamma}|\\
&=& \frac{\gamma}{c^{\gamma+1}}||\hat{\bbeta}_i^{�}|-|\hat{\bbeta}_j^{�}||
\quad \mbox{where} \quad c\in[\min(|\hat{\bbeta}_i^{�}|,|\hat{\bbeta}_j^{�}|),\max(|\hat{\bbeta}_i^{�}|,|\hat{\bbeta}_j^{�}|)] \\
&\le& \frac{\gamma}{\min(|\hat{\bbeta}_i^{�}|,|\hat{\bbeta}_j^{�}|)^{\gamma+1}}|\hat{\bbeta}_i^{�}-\hat{\bbeta}_j^{�}|.
\end{eqnarray*}
Then the equation (\ref{eq9}) implies that
\begin{eqnarray}\label{eqn1}
\mid\sum_{k=1}^{p} (q_{ik}
-q_{jk})\hat{\bbeta}_{k}\mid &\leq&
\frac{1}{\lambda_{2}}\|\hat{\mathbf{r}}(\lambda_{1}^{*},
\lambda_{2})\|\|\mathbf{x}_{1} - \mathbf{x}_{2}\| + \frac{\gamma\lambda_{1}^{*}}{\min(|\hat{\bbeta}_i^{�}|,|\hat{\bbeta}_j^{�}|)^{\gamma+1}}|\hat{\bbeta}_i^{�}-\hat{\bbeta}_j^{�}| \nonumber \\
\end{eqnarray}
dividing by $\|\mathbf{y}\|_{2}$, we obtain
\begin{eqnarray}\label{rev1eqn1}
\frac{1}{\|\mathbf{y}\|_{2}}\mid\sum_{k=1}^{p} (q_{ik}
-q_{jk})\hat{\bbeta}_{k}\mid
&\le& \frac{\sqrt{2(1 - \rho_{ij})}}{\lambda_{2}}
+\frac{\gamma\lambda_{1}^{*}}{\|\mathbf{y}\|_2\min(|\hat{\bbeta}_i^{�}|,|\hat{\bbeta}_j^{�}|)^{\gamma+1}}|\hat{\bbeta}_i^{�}-\hat{\bbeta}_j^{�}|
\end{eqnarray}
On the other hand, we have:
\begin{eqnarray}\label{eqnn2}
q_{ii} = - \sum_{s\neq i}^{} \frac{q_{is}}{\rho_{is}}
\hspace*{1cm} \mbox{and} \hspace*{1cm} q_{jj} = - \sum_{s\neq
j}^{} \frac{q_{js}}{\rho_{js}}.
\end{eqnarray}
Then
\begin{eqnarray}\label{eqnn3}
\sum_{k=1}^{p} (q_{ik}
-q_{jk})\hat{\bbeta}_{k} = \frac{-2}{1
- \rho_{ij}}[\hat{\bbeta}_{j} -\hat{\bbeta}_{i}] + 2SN
\end{eqnarray}
where
$
SN = \sum_{k\neq i,j}^{} \frac{1}{1 -
\rho^{2}_{ki}}[\hat{\bbeta}_{i} -
\rho_{ki}\hat{\bbeta}_{k}] + \frac{1}{1 -
\rho^{2}_{kj}}[\rho_{kj}\hat{\bbeta}_{k} -
\hat{\bbeta}_{j}].
$
If $\rho_{ki} = \rho_{kj} = \rho, \hspace*{3mm} \forall k = 1, ...,
p,$ then
$SN = \frac{p - 2}{1 - \rho^{2}}(\hat{\bbeta}_{i} - \hat{\bbeta}_{j}).$
So using (\ref{eqn1}) we have:
\begin{eqnarray*}\label{eqnnnnn5555555}
\frac{1}{\|\mathbf{y}\|_{2}}\left|\hat{\bbeta}_{j} -
\hat{\bbeta}_{i}\right| & \leq & \frac{1 - \rho^{2}}{2(p + \rho - 1)\lambda_{2}}
\left[\sqrt{2(1 - \rho)}
+\frac{\gamma\lambda_{1}^{*}}{\|\mathbf{y}\|_2\min(|\hat{\bbeta}_i^{�}|,|\hat{\bbeta}_j^{�}|)^{\gamma+1}}|\hat{\bbeta}_i^{�}-\hat{\bbeta}_j^{�}|\right]\\
\end{eqnarray*}
This completes the proof.
\end{proof}

\begin{proof} PROOF OF THEOREM \ref{mytheo1}: The proof is similar to that of Theorem 3.1
in Zou and Zhang (2009). We must only take account, in different inequalities, that

\begin{eqnarray*}\label{eqn3}
bn + \lambda_2 d & \leq &  \lambda_{\min}(\mathbf{\mathbf{X}^t\mathbf{X}) + \lambda_2\lambda_{\min}(\mathbf{Q}})\nonumber \\
& \leq &  \lambda_{\min}(\mathbf{\mathbf{X}^t\mathbf{X} +
\lambda_2\mathbf{Q}})\nonumber
\end{eqnarray*}
\begin{eqnarray*}
\lambda_{max}({\mathbf{X}^t\mathbf{X} + \boldsymbol{\lambda_2}\mathbf{Q}}) \nonumber
& \leq & \lambda_{max}(\mathbf{\mathbf{X}^t\mathbf{X}) + \lambda_{2}\lambda_{max}(\mathbf{Q}}) \nonumber \\
& \leq & Bn + \lambda_{2} d. \nonumber \\
\end{eqnarray*}
\end{proof}
\begin{proof} PROOF OF THEOREM \ref{mytheo2}: To prove this Theorem, we must show, as in
Zou \& Zhang (2009), that $(\mathbf{N}_{\mathcal{A}}\tilde{\boldsymbol{\bbeta}}_{\mathcal{A}}^{*}, 0)$ satisfies the
Karush-Kuhn-Tucker condition of (\ref{equ9}) with probability tending
to $1$. By the definition of $\tilde{\boldsymbol{\bbeta}}_{\mathcal{A}}^{*}$, it
suffices to show
\begin{eqnarray*}\label{eqn8}
 \mbox{Pr}(\forall j \in \mathcal{A}^{c}, |-2\mathbf{X}_{j}^t(\mathbf{y} - \mathbf{X}_{\mathcal{A}}\tilde{\boldsymbol{\bbeta}}_{\mathcal{A}}^{*}) + 2\sum_{k \in \mathcal{A}}q_{jk}\tilde{\bbeta}_{k}^{*}|\leq
 \lambda_1^{*}\hat{\omega}_{j})\rightarrow 1,
\end{eqnarray*}
or equivalently
\begin{eqnarray*}\label{eqn9}
 \mbox{Pr}(\exists j \in \mathcal{A}^{c}, |-2\mathbf{X}_{j}^t(\mathbf{y} - \mathbf{X}_{\mathcal{A}}\tilde{\boldsymbol{\bbeta}}_{\mathcal{A}}^{*}) + 2\sum_{k \in \mathcal{A}}q_{jk}\tilde{\bbeta}_{k}^{*}|>
 \lambda_1^{*}\hat{\omega}_{j})\rightarrow 0.
\end{eqnarray*}
The term $2\sum_{k \in \mathcal{A}}q_{jk}\tilde{\bbeta}_{k}^{*}$
does not appear in the proof of the same Theorem 4.2 for adaptive
elastic net in Zhang \& Zou (2009). So, taking into account this
difference, some modifications of the proof of Zou \& Zhang's
Theorem 4.2 are necessary. Let $\eta = \min_{k \in
\mathcal{A}}(|\bbeta_{k}^{*}|)$ and $\hat{\eta} = \min_{k \in
\mathcal{A}}(|\hat{\bbeta}(\mbox{Gril})_{k}^{*}|)$. We note that
\begin{eqnarray*}\label{eqn10}
 &  & \mbox{Pr}(\exists j \in \mathcal{A}^{c}, |-2\mathbf{X}_{j}^t(\mathbf{y} - \mathbf{X}_{\mathcal{A}}\tilde{\boldsymbol{\bbeta}}_{\mathcal{A}}^{*}) + 2\sum_{k \in \mathcal{A}}q_{jk}\tilde{\bbeta}_{k}^{*}|>
 \lambda_1^{*}\hat{\omega}_{j}) \nonumber \\
 & \leq & \sum_{j \in \mathcal{A}^{c}}\mbox{Pr}(|-2\mathbf{X}_{j}^t(\mathbf{y} - \mathbf{X}_{\mathcal{A}}\tilde{\boldsymbol{\bbeta}}_{\mathcal{A}}^{*}) + 2\sum_{k \in \mathcal{A}}q_{jk}\tilde{\bbeta}_{k}^{*}|>
 \lambda_1^{*}\hat{\omega}_{j}, \hat{\eta} > \eta/2) + \mbox{Pr}(\hat{\eta} \leq \eta/2).\nonumber
\end{eqnarray*}
Since
\begin{eqnarray}\label{eclair.eq1}
|\hat{\bbeta}(\mbox{Gril})_{j}| & \geq &  \eta - \|\boldsymbol{\bbeta^{*}} - \hat{\bbeta}(\mbox{Gril})\|_{2} \hspace*{3mm} \mbox{for all} \hspace*{3mm} j \in \mathcal{A},\nonumber\\
\end{eqnarray}
we have
\begin{eqnarray}\label{eclair.eq2}
\hat{\eta} & \geq &  \eta - \|\boldsymbol{\bbeta^{*}} - \hat{\bbeta}(\mbox{Gril})\|_{2}.
\end{eqnarray}
If $\hat{\eta} \leq \eta/2$, we have $\|\boldsymbol{\bbeta^{*}} - \hat{\bbeta}(\mbox{Gril})\|_{2} \geq \eta/2$ and so
$$\mbox{Pr}(\hat{\eta} \leq \eta/2) \leq \mbox{Pr}(\|\hat{\boldsymbol{\bbeta}}(\mbox{Gril}) - \boldsymbol{\bbeta}^{*}\|_{2} \geq \eta/2) \leq \frac{I\!\!E(\|\hat{\boldsymbol{\bbeta}}(\mbox{Gril}) - \boldsymbol{\bbeta}^{*}\|_{2}^{2})}{\eta^{2}/4}.$$
Then from Theorem \ref{mytheo1} we have
\begin{eqnarray}\label{eqn11}
\mbox{Pr}(\hat{\eta} \leq \eta/2) \leq
16\frac{\lambda_2^{2}D^{2}\|\boldsymbol{\bbeta}^{*}\|_{2}^{2} + Bpn\sigma^{2} +
p\lambda_1^{2}}{(bn + \lambda_2 d)^{2}\eta^{2}}.
\end{eqnarray}
Moreover, let $M = \left(\lambda_1^{*}/n\right)^{1/(1 +
\gamma)}$ and using similar arguments as in the proof of equation
(6.8) in Zou \& Zhang (2009), we have
\begin{eqnarray}\label{eqn12}
 &  & \sum_{j \in \mathcal{A}^{c}}\mbox{Pr}(|-2\mathbf{X}_{j}^t(\mathbf{y} - \mathbf{X}_{\mathcal{A}}\tilde{\boldsymbol{\bbeta}}_{\mathcal{A}}^{*}) + 2\sum_{k \in \mathcal{A}}q_{jk}\tilde{\bbeta}_{k}^{*}|>
 \lambda_1^{*}\hat{\omega}_{j}, \hat{\eta} > \eta/2) \nonumber \\
 & \leq & \sum_{j \in \mathcal{A}^{c}}\mbox{Pr}(|-2\mathbf{X}_{j}^t(\mathbf{y} - \mathbf{X}_{\mathcal{A}}\tilde{\boldsymbol{\bbeta}}_{\mathcal{A}}^{*}) + 2\sum_{k \in \mathcal{A}}q_{jk}\tilde{\bbeta}_{k}^{*}|>
 \lambda_1^{*}\hat{\omega}_{j}, \hat{\eta} > \eta/2, |\hat{\bbeta}(\mbox{Gril})_{j}| \leq M) \nonumber \\
&  & + \sum_{j \in \mathcal{A}^{c}}\mbox{Pr}(|\hat{\bbeta}(\mbox{Gril})_{j}| > M)\nonumber \\
& \leq & \frac{4M^{2\gamma}}{\lambda_{1}^{*2}}I\!\!E\left(\sum_{j \in
\mathcal{A}^{c}}|-\mathbf{X}_{j}^t(\mathbf{y} -
\mathbf{X}_{\mathcal{A}}\tilde{\boldsymbol{\bbeta}}_{\mathcal{A}}^{*}) + \sum_{k
\in \mathcal{A}}q_{jk}\tilde{\bbeta}_{k}^{*}|^{2}I({\hat{\eta} > \eta/2})\right) \nonumber \\
&  & + \frac{I\!\!E\left(\|\hat{\boldsymbol{\bbeta}}(\mbox{Gril}) - \boldsymbol{\bbeta}^{*}\|_{2}^{2}\right)}{M^{2}}\nonumber \\
& \leq & \frac{4M^{2\gamma}}{\lambda_1^{*2}}I\!\!E\left(\sum_{j \in
\mathcal{A}^{c}}|-\mathbf{X}_{j}^t(\mathbf{y} -
\mathbf{X}_{\mathcal{A}}\tilde{\boldsymbol{\bbeta}}_{\mathcal{A}}^{*}) + \sum_{k
\in \mathcal{A}}q_{jk}\tilde{\bbeta}_{k}^{*}|^{2}I({\hat{\eta} > \eta/2})\right) \nonumber \\
&  & + 4\frac{\lambda_2^{2}D^{2}\|\boldsymbol{\bbeta}^{*}\|_{2}^{2} + Bpn\sigma^{2} +
p\lambda_1^{2}}{(bn + \lambda_2 d)^{2}M^{2}}.\nonumber \\
\end{eqnarray}
We have used the result of Theorem \ref{mytheo1} for the second term of the
last inequality. On the other hand, it is easy to show that
\begin{eqnarray*}\label{eqn13}
\sum_{j \in \mathcal{A}^{c}}|-\mathbf{X}_{j}^t(\mathbf{y} -
\mathbf{X}_{\mathcal{A}}\tilde{\boldsymbol{\bbeta}}_{\mathcal{A}}^{*}) + \sum_{k
\in \mathcal{A}}q_{jk}\tilde{\bbeta}_{k}^{*}|^{2} & \leq & 2\sum_{j
\in \mathcal{A}^{c}}|\mathbf{X}_{j}^t(\mathbf{y} -
\mathbf{X}_{\mathcal{A}}\tilde{\boldsymbol{\bbeta}}_{\mathcal{A}}^{*})|^{2} +
2\sum_{j \in \mathcal{A}^{c}}\left(\sum_{k \in
\mathcal{A}}q_{jk}\tilde{\bbeta}_{k}^{*}\right)^{2}.\nonumber
\end{eqnarray*}
From Zou \& Zhang (2009), in page $16$, we have
\begin{eqnarray*}\label{eqn14}
\sum_{j \in \mathcal{A}^{c}}|\mathbf{X}_{j}^t(\mathbf{y} -
\mathbf{X}_{\mathcal{A}}\tilde{\boldsymbol{\bbeta}}_{\mathcal{A}}^{*})|^{2}
& \leq & 2Bn.Bn\|\boldsymbol{\bbeta}_{\mathcal{A}}^{*} -
\tilde{\boldsymbol{\bbeta}}_{\mathcal{A}}^{*}\|_{2}^{2} +
2Bn\|\mathbf{\varepsilon}\|_{2}^{2},
\end{eqnarray*}
which leads to the following inequality
\begin{eqnarray}\label{eqn15}
&  & I\!\!E\left(\sum_{j \in
\mathcal{A}^{c}}|\mathbf{X}_{j}^t(\mathbf{y} -
\mathbf{X}_{\mathcal{A}}\tilde{\boldsymbol{\bbeta}}_{\mathcal{A}}^{*})|^{2}I({\hat{\eta} > \eta/2})\right) \nonumber \\
 & \leq & 2B^{2}n^{2}I\!\!E\left(\|\boldsymbol{\bbeta}_{\mathcal{A}}^{*} - \tilde{\boldsymbol{\bbeta}}_{\mathcal{A}}^{*}\|_{2}^{2}I({\hat{\eta} > \eta/2})\right) + 2Bnp\sigma^{2}.\nonumber\\
\end{eqnarray}
We now bound $I\!\!E\left(\|\boldsymbol{\bbeta}_{\mathcal{A}}^{*} -
\tilde{\boldsymbol{\bbeta}}_{\mathcal{A}}^{*}\|_{2}^{2}I({\hat{\eta} >
\eta/2})\right).$ Let
$$\tilde{\boldsymbol{\bbeta}}_{\mathcal{A}}^{*}(\lambda_2, 0) = \arg\min_{\boldsymbol{\bbeta}}\left\{\|\mathbf{y} - \mathbf{X}_{\mathcal{A}}\boldsymbol{\bbeta}\|_{2}^{2} + \lambda_2\mathbf{\bbeta}^t\mathbf{Q}_{\mathcal{A}}\mathbf{\bbeta}\right\}.$$
Then, as in Zou \& Zhang (2009), by using the same arguments for
deriving (\ref{eqn3}), we have
\begin{eqnarray}\label{eqn16}
\|\tilde{\boldsymbol{\bbeta}}_{\mathcal{A}}^{*} -
\tilde{\boldsymbol{\bbeta}}_{\mathcal{A}}^{*}(\lambda_2, 0)\|_{2} \leq
\frac{\lambda_1^{*}.\max_{j \in
{\mathcal{A}}}\hat{\omega}_{j}\sqrt{|\mathcal{A}|}}{\lambda_{\min}(\mathbf{X}_{\mathcal{A}}^t\mathbf{X}_{\mathcal{A}})
+ \lambda_2 d} \leq \frac{\lambda_1^{*}\hat{\eta}^{-\gamma}\sqrt{p}}{bn +
\lambda_2 d}.
\end{eqnarray}
Following the similar arguments used in the proof of Theorem \ref{mytheo1}, we
obtain
\begin{eqnarray}\label{eqn17}
&  & I\!\!E\left(\|\boldsymbol{\bbeta}_{\mathcal{A}}^{*} -
\tilde{\boldsymbol{\bbeta}}_{\mathcal{A}}^{*}\|_{2}^{2}I({\hat{\eta} >
\eta/2})\right)\nonumber \\
& \leq & 4\frac{\lambda_2^{2}D^{2}\|\boldsymbol{\bbeta}^{*}\|_{2}^{2} + Bpn\sigma^{2} +
\lambda_1^{*2}(\eta/2)^{-2\gamma}p}{(bn + \lambda_2 d)^{2}}. \nonumber
\end{eqnarray}
Now, we bound the second term $ \sum_{j \in
\mathcal{A}^{c}}\left(\sum_{k \in
\mathcal{A}}q_{jk}\tilde{\bbeta}_{k}^{*}\right)^{2}$. In fact, we
have
\begin{eqnarray}\label{eqn18}
\sum_{j \in \mathcal{A}^{c}}\left(\sum_{k \in \mathcal{A}}q_{jk}\tilde{\bbeta}_{k}^{*}\right)^{2} & \leq & \sum_{j \in \mathcal{A}^{c}}\left[\left(\sum_{k \in \mathcal{A}}q_{jk}^{2}\right)\left(\sum_{k \in \mathcal{A}}\tilde{\bbeta}_{k}^{*2}\right)\right] \left(\mbox{Cauchy-Schwarz inequality}\right)\nonumber \\
& \leq & \sum_{j \in \mathcal{A}^{c}}\left[\left(\sum_{k \in \mathcal{A}}q_{jk}^{2}\right)\|\tilde{\boldsymbol{\bbeta}}_{\mathcal{A}}^{*}\|_{2}^{2}\right] \nonumber \\
& \leq & \|\tilde{\boldsymbol{\bbeta}}_{\mathcal{A}}^{*}\|_{2}^{2}\sum_{j=1}^{p}\sum_{k=1}^{p}q_{jk}^{2} \nonumber \\
& = &
\|\tilde{\boldsymbol{\bbeta}}_{\mathcal{A}}^{*}\|_{2}^{2}.\|\mathbf{Q}\|_{F}^{2} \hspace*{5mm}(\|.\|_{F} \hspace*{1mm}\mbox{is the Frobenius norm})\nonumber \\
& \leq &
p\|\tilde{\boldsymbol{\bbeta}}_{\mathcal{A}}^{*}\|_{2}^{2}.\|\mathbf{Q}\|_{2}^{2} \hspace*{5mm}(\|.\|_{2} \hspace*{1mm}\mbox{is the spectral norm})\nonumber \\
& \leq & pD^{2}\|\tilde{\boldsymbol{\bbeta}}_{\mathcal{A}}^{*}\|_{2}^{2}\nonumber \\
& \leq & 2pD^{2}\|\tilde{\boldsymbol{\bbeta}}_{\mathcal{A}}^{*} - \boldsymbol{\bbeta}_{\mathcal{A}}^{*}\|_{2}^{2} + 2pD^{2}\|\boldsymbol{\bbeta}_{\mathcal{A}}^{*}\|_{2}^{2}.\nonumber \\
\end{eqnarray}
It leads to the following inequality
\begin{eqnarray}\label{eqn19}
&  & I\!\!E\left(\sum_{j \in \mathcal{A}^{c}}\left(\sum_{k \in \mathcal{A}}q_{jk}\tilde{\bbeta}_{k}^{*}\right)^{2}I({\hat{\eta} > \eta/2})\right) \nonumber \\
 & \leq & 2pD^{2}I\!\!E\left(\|\boldsymbol{\bbeta}_{\mathcal{A}}^{*} - \tilde{\boldsymbol{\bbeta}}_{\mathcal{A}}^{*}\|_{2}^{2}I({\hat{\eta} > \eta/2})\right) + 2pD^{2}\|\boldsymbol{\bbeta}_{\mathcal{A}}^{*}\|_{2}^{2}.\nonumber
\end{eqnarray}
The combination of (\ref{eqn11}), (\ref{eqn12}), (\ref{eqn15}) and (\ref{eqn19}) yields
\begin{eqnarray*}\label{eqn20}
 &  & \mbox{Pr}(\exists j \in \mathcal{A}^{c}, |-2\mathbf{X}_{j}^t(\mathbf{y} - \mathbf{X}_{\mathcal{A}}\tilde{\boldsymbol{\bbeta}}_{\mathcal{A}}^{*}) + 2\sum_{k \in \mathcal{A}}q_{jk}\tilde{\bbeta}_{k}^{*}|>
 \lambda_1^{*}\hat{\omega}_{j}) \nonumber \\
 & \leq & \frac{4M^{2\gamma}n}{\lambda_1^{*2}}\left(16(B^{2}n + D^{2}p)\frac{\lambda_2^{2}D^{2}\|\boldsymbol{\bbeta}^{*}\|_{2}^{2} +
Bpn\sigma^{2} + \lambda_1^{*2}(\eta/2)^{-2\gamma}p}{(bn
+ \lambda_2 d)^{2}} + 4Bp\sigma^{2}\right)\nonumber \\
&  & + 16D^{2}\frac{M^{2\gamma}p}{\lambda_1^{*2}}\|\boldsymbol{\bbeta}^{*}\|_{2}^{2}\nonumber \\
&  & + \frac{\lambda_2^{2}D^{2}\|\boldsymbol{\bbeta}^{*}\|_{2}^{2} + Bpn\sigma^{2} +
\lambda_1^{2}p}{(bn
+ \lambda_2 d)^{2}}\frac{4}{M^{2}}\nonumber \\
&  & + \frac{\lambda_2^{2}D^{2}\|\boldsymbol{\bbeta}^{*}\|_{2}^{2} + Bpn\sigma^{2} +
\lambda_1^{2}p}{(bn
+ \lambda_2 d)^{2}}\frac{16}{\eta^{2}}\nonumber \\
&  & \equiv L_{1} + L_{2} + L_{3} + L_{4}.
\end{eqnarray*}
We have chosen $\gamma > 2\nu/(1 - \nu)$, then under
conditions (C1)-(C6) it follows that
\begin{eqnarray}\label{eqn21}
L_{1} & = & O\left(\left(\frac{\lambda^{*}}{\sqrt{n}}n^{\frac{(1 + \gamma)(1 - \nu) - 1}{2}}\right)^{-\frac{2}{1 + \gamma}}\right)\rightarrow 0, \nonumber \\
L_{2} & = & O\left(\left(\frac{\lambda_{1}^{*}}{n}\right)^{\frac{2\gamma}{1 + \gamma}}\frac{p\|\bbeta^{*}\|_{2}^{2}}{\lambda_{1}^{*2}}\right)
\rightarrow 0, \nonumber \\
L_{3} & = & O\left(\frac{p}{n}\left(\frac{n}{\lambda_{1}^{*}}\right)^{\frac{2}{1 + \gamma}}\right)\rightarrow 0, \nonumber \\
L_{4} & = & O\left(\frac{p}{n}\frac{1}{\eta^{2}}\right) O\left(\left(\lambda_{1}^{*}\sqrt{\frac{p}{n}}\eta^{-\gamma}\right)^{2/\gamma}\left(\frac{p}{n}
\left(\frac{n}{\lambda_{1}^{*}}\right)^{2/(2 + \gamma)}\right)^{(1 + \gamma)/\gamma}p^{-2/\gamma}\right)\rightarrow 0. \nonumber \\
\end{eqnarray}
\end{proof}
\begin{proof} PROOF OF THEOREM \ref{mytheo3}: The proof for selection consistency of the AdaGril is exactly similar to the proof of selection consistency of AdaEnet (cf. pages 1748-1749 in Zou and Zhang 2009).

We now prove the asymptotic normality. For
convenience we put
\textcolor{black}{\begin{eqnarray*}\label{eqn22}
z_{n} = \alpha^t\Sigma_{\mathcal{A}}^{\frac{1}{2}}\left(I +
\lambda_2\Sigma_{\mathcal{A}}^{-1}\mathbf{Q}_{\mathcal{A}}\right)\mathbf{N}_{\mathcal{A}}^{-1}\left(\hat{\bbeta}(\mbox{AdaGril})_{\mathcal{A}}-
\boldsymbol{\bbeta}_{\mathcal{A}}^{*}\right).
\end{eqnarray*}}
Note that
\textcolor{black}{\begin{eqnarray*}\label{eqn23}
&  & \alpha^t\Sigma_{\mathcal{A}}^{\frac{1}{2}}\left(I + \lambda_2\Sigma_{\mathcal{A}}^{-1}\mathbf{Q}_{\mathcal{A}}\right)\left(\tilde{\boldsymbol{\bbeta}}_{\mathcal{A}}^{*}-  \mathbf{N}_{\mathcal{A}}^{-1}\boldsymbol{\bbeta}_{\mathcal{A}}^{*}\right)\nonumber \\
& = & \alpha^t\Sigma_{\mathcal{A}}^{\frac{1}{2}}\left(I + \lambda_2\Sigma_{\mathcal{A}}^{-1}\mathbf{Q}_{\mathcal{A}}\right)\left(\tilde{\boldsymbol{\bbeta}}_{\mathcal{A}}^{*}-  \mathbf{N}_{\mathcal{A}}^{-1}\boldsymbol{\bbeta}_{\mathcal{A}}^{*} + \tilde{\boldsymbol{\bbeta}}_{\mathcal{A}}^{*}(\lambda_2, 0) - \tilde{\boldsymbol{\bbeta}}_{\mathcal{A}}^{*}(\lambda_2, 0) + \boldsymbol{\bbeta}_{\mathcal{A}}^{*} - \boldsymbol{\bbeta}_{\mathcal{A}}^{*}\right)\nonumber \\
 & = & \alpha^t\Sigma_{\mathcal{A}}^{\frac{1}{2}}\left(I + \lambda_2\Sigma_{\mathcal{A}}^{-1}\mathbf{Q}_{\mathcal{A}}\right)\left(\boldsymbol{\bbeta}_{\mathcal{A}}^{*} - \mathbf{N}_{\mathcal{A}}^{-1}\boldsymbol{\bbeta}_{\mathcal{A}}^{*}\right)+ \alpha^t\Sigma_{\mathcal{A}}^{\frac{1}{2}}\left(I + \lambda_2\Sigma_{\mathcal{A}}^{-1}\mathbf{Q}_{\mathcal{A}}\right)\left(\tilde{\boldsymbol{\bbeta}}_{\mathcal{A}}^{*}-  \tilde{\boldsymbol{\bbeta}}_{\mathcal{A}}^{*}(\lambda_2, 0)\right)\nonumber \\
 &  & +\hspace*{2mm} \alpha^t\Sigma_{\mathcal{A}}^{\frac{1}{2}}\left(I + \lambda_2\Sigma_{\mathcal{A}}^{-1}\mathbf{Q}_{\mathcal{A}}\right)\left(\tilde{\boldsymbol{\bbeta}}_{\mathcal{A}}^{*}(\lambda_2, 0) - \boldsymbol{\bbeta}_{\mathcal{A}}^{*}\right).\nonumber \\
\end{eqnarray*}}
In addition, we have
\begin{eqnarray*}\label{eqn24}
\Sigma_{\mathcal{A}}^{\frac{1}{2}}\left(I +
\lambda_2\Sigma_{\mathcal{A}}^{-1}\mathbf{Q}_{\mathcal{A}}\right)\left(\tilde{\boldsymbol{\bbeta}}_{\mathcal{A}}^{*}(\lambda_2,
0) - \boldsymbol{\bbeta}_{\mathcal{A}}^{*}\right) =
-\lambda_2\Sigma_{\mathcal{A}}^{-\frac{1}{2}}\mathbf{Q}_{\mathcal{A}}\boldsymbol{\bbeta}_{\mathcal{A}}^{*}
+
\Sigma_{\mathcal{A}}^{-\frac{1}{2}}\mathbf{X}_{\mathcal{A}}^t\varepsilon.
\end{eqnarray*}
Therefore, by Theorem \ref{mytheo2} it follows that with probability tending
to $1$, $z_{n} = T_{1} + T_{2} + T_{3}$, where

\hspace{1.5cm}\textcolor{black}{$T_{1}  =  \alpha^t\Sigma_{\mathcal{A}}^{\frac{1}{2}}\left(I + \lambda_2\Sigma_{\mathcal{A}}^{-1}\mathbf{Q}_{\mathcal{A}}\right)\mathbf{K}_{\mathcal{A}}\boldsymbol{\bbeta}_{\mathcal{A}}^{*} -
\alpha^t\lambda_2\Sigma_{\mathcal{A}}^{-\frac{1}{2}}\mathbf{Q}_{\mathcal{A}}\boldsymbol{\bbeta}_{\mathcal{A}}^{*}
$}, where

 \textcolor{black}{$$\mathbf{K}_{\mathcal{A}} = \mathbf{I}_{\mathcal{A}} - \mathbf{N}_{\mathcal{A}}^{-1} = \mbox{diag}\left(\frac{\lambda_{2}q_{i}}{n + \lambda_{2}q_{i}}\right)_{i \in \mathcal{A}}, $$}


\hspace{1.5cm}$T_{2}  =  \alpha^t\Sigma_{\mathcal{A}}^{\frac{1}{2}}\left(I +
\lambda_2\Sigma_{\mathcal{A}}^{-1}\mathbf{Q}_{\mathcal{A}}\right)\left(\tilde{\boldsymbol{\bbeta}}_{\mathcal{A}}^{*}-
\tilde{\boldsymbol{\bbeta}}_{\mathcal{A}}^{*}(\lambda_2, 0)\right)$,

\hspace{1.5cm}$T_{3} =
\alpha^t\Sigma_{\mathcal{A}}^{-\frac{1}{2}}\mathbf{X}_{\mathcal{A}}^t\varepsilon
$.

We now show that $T_{1} = o(1)$, $T_{2} = o_{P}(1)$ and
$T_{3}\rightarrow_{d} N(0, \sigma^{2})$. Then by the Slusky's
theorem we know $z_{n}\rightarrow_{d}N(0, \sigma^{2})$. From (C1)
and the fact that $\alpha^t\alpha = 1$, we have
\textcolor{black}{\begin{eqnarray*}\label{eqn25}
T_{1}^{2} & \leq & 2\|\Sigma_{\mathcal{A}}^{\frac{1}{2}}\left(I + \lambda_2\Sigma_{\mathcal{A}}^{-1}\mathbf{Q}_{\mathcal{A}}\right)\mathbf{K}_{\mathcal{A}}\boldsymbol{\bbeta}_{\mathcal{A}}^{*}\|_{2}^{2}  + 2\|\lambda_2\Sigma_{\mathcal{A}}^{-\frac{1}{2}}\mathbf{Q}_{\mathcal{A}}\boldsymbol{\bbeta}_{\mathcal{A}}^{*}\|_{2}^{2}\nonumber \\
& \leq & 2\|\mathbf{K}_{\mathcal{A}}\|_{2}^{2}\|\Sigma_{\mathcal{A}}^{\frac{1}{2}}\left(I + \lambda_2\Sigma_{\mathcal{A}}^{-1}\mathbf{Q}_{\mathcal{A}}\right)\|_{2}^{2}\|\boldsymbol{\bbeta}_{\mathcal{A}}^{*}\|_{2}^{2} + 2\lambda_{2}^{2}\|\Sigma_{\mathcal{A}}^{-\frac{1}{2}}\mathbf{Q}_{\mathcal{A}}\|_{2}^{2}\|\boldsymbol{\bbeta}_{\mathcal{A}}^{*}\|_{2}^{2}\nonumber \\
& \leq & 2\max_{i \in \mathcal{A}}\left(\frac{\lambda_{2}q_{i}}{n + \lambda_{2}q_{i}}\right)^{2}\|\Sigma_{\mathcal{A}}^{\frac{1}{2}}\|_{2}^{2}\|\left(I + \lambda_2\Sigma_{\mathcal{A}}^{-1}\mathbf{Q}_{\mathcal{A}}\right)\|_{2}^{2}\|\boldsymbol{\bbeta}_{\mathcal{A}}^{*}\|_{2}^{2} + 2\lambda_{2}^{2}\|\Sigma_{\mathcal{A}}^{-\frac{1}{2}}\|_{2}^{2}\|\mathbf{Q}_{\mathcal{A}}\|_{2}^{2}\|\boldsymbol{\bbeta}_{\mathcal{A}}^{*}\|_{2}^{2}\nonumber \\
& \leq & 2\max_{i \in \mathcal{A}}\left(\frac{\lambda_{2}q_{i}}{n + \lambda_{2}q_{i}}\right)^{2}\|\Sigma_{\mathcal{A}}^{\frac{1}{2}}\|_{2}^{2}\left(\|I\|_{2} + \lambda_2\|\Sigma_{\mathcal{A}}^{-1}\|_{2}\|\mathbf{Q}_{\mathcal{A}}\|_{2}\right)^{2}\|\boldsymbol{\bbeta}_{\mathcal{A}}^{*}\|_{2}^{2} + 2\lambda_{2}^{2}\|\Sigma_{\mathcal{A}}^{-\frac{1}{2}}\|_{2}^{2}\|\mathbf{Q}_{\mathcal{A}}\|_{2}^{2}\|\boldsymbol{\bbeta}_{\mathcal{A}}^{*}\|_{2}^{2}\nonumber \\
& \leq & 2\max_{i \in \mathcal{A}}\left(\frac{\lambda_{2}q_{i}}{n + \lambda_{2}q_{i}}\right)^{2}Bn\left(1 + \frac{\lambda_{2}D}{bn}\right)^{2}\|\boldsymbol{\bbeta}_{\mathcal{A}}^{*}\|_{2}^{2} + 2\lambda_{2}^{2}\frac{1}{bn}D^{2}\|\boldsymbol{\bbeta}_{\mathcal{A}}^{*}\|_{2}^{2}.\nonumber \\
\end{eqnarray*}}
To obtain previous inequalities we have used sub-multiplicativity and consistency of the $\|.\|_{2}$ matrix norms. Hence it follows from (C5) and (C6) that $T_{1} = o(1)$. Similarly, we can
bound $T_{2}$ as follows
\begin{eqnarray*}\label{eqn26}
T_{2}^{2} & \leq & \|\Sigma_{\mathcal{A}}^{\frac{1}{2}}\|_{2}^{2}\|\left(I +
\lambda_2\Sigma_{\mathcal{A}}^{-1}\mathbf{Q}_{\mathcal{A}}\right)\|_{2}^{2}\|\tilde{\boldsymbol{\bbeta}}_{\mathcal{A}}^{*}-
\tilde{\boldsymbol{\bbeta}}_{\mathcal{A}}^{*}(\lambda_2, 0)\|_{2}^{2} \nonumber \\
& \leq & Bn\left(1 + \frac{\lambda_{2}D}{bn}\right)^{2}\|\tilde{\boldsymbol{\bbeta}}_{\mathcal{A}}^{*} - \tilde{\boldsymbol{\bbeta}}_{\mathcal{A}}^{*}(\lambda_2, 0)\|_{2}^{2} \nonumber \\
& \leq & Bn\left(1 + \frac{\lambda_{2}D}{bn}\right)^{2}\left(\frac{\lambda_1^{*}\hat{\eta}^{-\gamma}p}{bn + \lambda_2 d}\right)^{2} \nonumber \\
\end{eqnarray*}
where we have used (\ref{eqn16}) in the last step. Then $T_{2}^{2} =
O_{\mbox{p}}(1)/n^{2}$.
Finally, following the same arguments used in Theorem 3.3 of Zou \&
Zhang (2009), we obtain that $T_{3}\rightarrow_{d}N(0, \sigma^{2})$.
This completes the proof.
\end{proof}
\begin{proof} PROOF OF THEOREM \ref{oracle.theo}:
Now, we consider the Adaptive Gril estimator, with Gril estimates as initial weights.
The adaptive Gril estimates are defined by
\begin{equation} \label{oracle1}
\hat{\bbeta}(\mbox{AdaGril}) =\arg\min_{\boldsymbol{\boldsymbol{\bbeta}}} \textcolor{black}{\mathbf{N}}\left\{\|\mathbf{y} -
\mathbf{X}\boldsymbol{\boldsymbol{\bbeta}}\|_{2}^{2} + \lambda_2\boldsymbol{\boldsymbol{\bbeta}}^t\mathbf{Q}
\boldsymbol{\boldsymbol{\bbeta}} +
\lambda_1^{*}\sum_{j=1}^{p} \hat{\omega}_{j}|\boldsymbol{\bbeta}_{j}|\right\},
\end{equation}
where
\begin{equation} \label{poids}
\hat{\omega}_{j} = \max(\frac{1}{|\hat{\bbeta}(\mbox{Gril})_{j}|}, 1).
\end{equation}
Then, the minimizer of (\ref{oracle1}) is also the minimizer of the Adaptive Lasso problem on augmented data  $(\tilde{\mathbf{y}},\mathbf{\tilde{X}})$ defined in (\ref{augdata}). So, we have
\begin{eqnarray*}\label{weqn1}
\|\tilde{\mathbf{y}} - \mathbf{\tilde{X}}\hat{\bbeta}(\mbox{AdaGril})\|_{2}^{2} + \lambda_1^{*}\sum_{j=1}^{p} \hat{\omega}_{j}|\hat{\bbeta}(\mbox{AdaGril})_{j}| & \leq &
\|\tilde{\mathbf{y}} -
\mathbf{\tilde{X}}\boldsymbol{\boldsymbol{\bbeta}^{*}}\|_{2}^{2} + \lambda_1^{*}\sum_{j=1}^{p} \hat{\omega}_{j}|\boldsymbol{\bbeta}^{*}_{j}|.
\end{eqnarray*}
Since $tilde{\mathbf{y}} = \mathbf{\tilde{X}}\boldsymbol{\bbeta}^{*} + \mathbf{\tilde{\varepsilon}}$, the latter is equivalent to
\begin{eqnarray*}\label{weqn3}
\|\mathbf{\tilde{X}}\boldsymbol{\bbeta^{*}} - \mathbf{\tilde{X}}\hat{\bbeta}(\mbox{AdaGril})\|_{2}^{2}  & \leq & \lambda_1^{*}\sum_{j=1}^{p} \hat{\omega}_{j}(|\boldsymbol{\bbeta}^{*}_{j}| - |\hat{\bbeta}(\mbox{AdaGril})_{j}|) + 2\mathbf{\tilde{\varepsilon}}^t\mathbf{\tilde{X}}(\boldsymbol{\bbeta^{*}} - \hat{\bbeta}(\mbox{AdaGril})).
\end{eqnarray*}
Using the definition of $\mathbf{\tilde{X}}$ and $\mathbf{\tilde{\varepsilon}}$ on the third term of the latter inequality, we have
\begin{eqnarray}\label{weqn5}
\|\mathbf{\tilde{X}}\boldsymbol{\bbeta^{*}} - \mathbf{\tilde{X}}\hat{\bbeta}(\mbox{AdaGril})\|_{2}^{2}  & \leq & \lambda_1^{*}\sum_{j=1}^{p} \hat{\omega}_{j}(|\bbeta^{*}_{j}| - |\hat{\bbeta}(\mbox{AdaGril})_{j}|) + 2\mathbf{\varepsilon}^t\mathbf{X}(\boldsymbol{\bbeta^{*}} - \hat{\bbeta}(\mbox{AdaGril}))\nonumber\\
&   & - 2\lambda_2\boldsymbol{\bbeta^{*}}^t\mathbf{Q}(\boldsymbol{\bbeta^{*}} - \hat{\bbeta}(\mbox{AdaGril})).
\end{eqnarray}
Obviously, we have
\begin{eqnarray}\label{weqn6}
\sum_{j=1}^{p} \hat{\omega}_{j}(|\bbeta^{*}_{j}| - |\hat{\bbeta}(\mbox{AdaGril})_{j}|)
& \leq & \sum_{j=1}^{p} \hat{\omega}_{j}|\bbeta^{*}_{j} - \hat{\bbeta}(\mbox{AdaGril})_{j}|
\end{eqnarray}
%
and
\begin{eqnarray}\label{weqn8}
- 2\lambda_2\boldsymbol{\bbeta^{*}}^t\mathbf{Q}(\boldsymbol{\bbeta^{*}} - \hat{\bbeta}(\mbox{AdaGril}))
& \leq & 2\lambda_2 \|\mathbf{Q}\boldsymbol{\bbeta^{*}}\|_{\infty} \|\boldsymbol{\bbeta^{*}} - \hat{\bbeta}(\mbox{AdaGril})\|_{1}.
\end{eqnarray}
For $0 <\theta \leq 1$, consider the set $\Gamma = \{ \max_{j=1, ..., p} 2|U_{j}| \leq \theta\lambda_{1}\}$ with $U_{j} = n^{-1} \sum_{i=1}^{n} x_{i,j}\varepsilon_{i}$.

Therefore on the set $\Gamma$ and using (\ref{weqn5}), (\ref{weqn6}) and (\ref{weqn8}), we obtain
\begin{eqnarray}\label{weqn9}
\|\mathbf{\tilde{X}}\boldsymbol{\bbeta^{*}} - \mathbf{\tilde{X}}\hat{\bbeta}(\mbox{AdaGril})\|_{2}^{2}  & \leq & \lambda_1^{*}\sum_{j=1}^{p} \hat{\omega}_{j}(|\bbeta^{*}_{j}| - |\hat{\bbeta}(\mbox{AdaGril})_{j}|) + \theta\lambda_1^{*}\|\boldsymbol{\bbeta^{*}} - \hat{\bbeta}(\mbox{AdaGril})\|_{1}\nonumber\\
& +  & 2\lambda_2\|\mathbf{Q}^t\boldsymbol{\bbeta^{*}}\|_{\infty} \|\boldsymbol{\bbeta^{*}} - \hat{\bbeta}(\mbox{AdaGril})\|_{1}.
\end{eqnarray}
Tacking $\theta = \frac{1}{4}$ and $\lambda_{2} = \frac{\lambda_1^{*}}{8\|\mathbf{Q}\boldsymbol{\bbeta^{*}}\|_{\infty}}$ and adding $2^{-1}\lambda_1^{*}\|\boldsymbol{\bbeta^{*}} - \hat{\bbeta}(\mbox{AdaGril})\|_{1}$ to both sides of the previous inequality, we have
\begin{eqnarray}\label{weqn10}
\|\mathbf{\tilde{X}}\boldsymbol{\bbeta^{*}} - \mathbf{\tilde{X}}\hat{\bbeta}(\mbox{AdaGril})\|_{2}^{2} + \frac{\lambda_1^{*}}{2}\|\boldsymbol{\bbeta^{*}} - \hat{\bbeta}(\mbox{AdaGril})\|_{1} & \leq & \lambda_1^{*}\sum_{j=1}^{p} \hat{\omega}_{j}(|\bbeta^{*}_{j}| - |\hat{\bbeta}(\mbox{AdaGril})_{j}|)\nonumber\\
& + & \lambda_1^{*}\sum_{j=1}^{p}|\bbeta^{*}_{j} - \hat{\bbeta}(\mbox{AdaGril})_{j}|
\end{eqnarray}
 Since
$\hat{\omega}_{j} = \max\left(1/|\hat{\bbeta}(\mbox{Gril})_{j}|, 1\right)$ for all $j = 1, ..., p,$ we have
\begin{eqnarray}\label{wweqn10}
\|\mathbf{\tilde{X}}\boldsymbol{\bbeta^{*}} - \mathbf{\tilde{X}}\hat{\bbeta}(\mbox{AdaGril})\|_{2}^{2} + \frac{\lambda_1^{*}}{2}\|\boldsymbol{\bbeta^{*}} - \hat{\bbeta}(\mbox{AdaGril})\|_{1} & \leq & \lambda_1^{*}\sum_{j=1}^{p} \hat{\omega}_{j}(|\bbeta^{*}_{j}| - |\hat{\bbeta}(\mbox{AdaGril})_{j}|)\nonumber\\
& + & \lambda_1^{*}\sum_{j=1}^{p} \hat{\omega}_{j}|\bbeta^{*}_{j} - \hat{\bbeta}(\mbox{AdaGril})_{j}|
\end{eqnarray}
Let $\delta = 8\psi^{-1}\lambda_1^{*}s,\eta = \min_{j \in \mathcal{A}}(|\bbeta_{j}^{*}|),  \omega_{\max}(\mathcal{A}) = \max_{j \in \mathcal{A}}\hat{\omega}_{j}.$
Hebiri and Van De Geer (2010) show that on $\Gamma$
\begin{eqnarray}\label{revweqn17}
\delta & \geq &  \|\boldsymbol{\bbeta^{*}} - \hat{\bbeta}(\mbox{Gril})\|_{\infty}.\nonumber
\end{eqnarray}
Suppose now that $\eta \geq 2\delta,$ then we have
$\eta \geq 2\delta \geq 2\|\boldsymbol{\bbeta^{*}} - \hat{\bbeta}(\mbox{Gril})\|_{\infty},$ and
\begin{eqnarray}\label{weqn17}
|\hat{\bbeta}(\mbox{Gril})_{j}| & \geq &  \eta - \|\boldsymbol{\bbeta^{*}} - \hat{\bbeta}(\mbox{Gril})\|_{\infty}\nonumber\\
& \geq & \frac{\eta}{2} \hspace*{3mm} \mbox{for all} \hspace*{3mm} j \in \mathcal{A},
\end{eqnarray}
Hence, we deduce that
\begin{eqnarray}\label{weqn18}
\omega_{\max}(\mathcal{A}) & \leq &  \max\left\{\frac{2}{\eta}, 1\right\}.
\end{eqnarray}
Using the fact that $|\bbeta^{*}_{j}| - |\hat{\bbeta}(\mbox{AdaGril})_{j}| + |\bbeta^{*}_{j} - \hat{\bbeta}(\mbox{AdaGril})_{j}| = 0$, for all $j \in \mathcal{A}^{c}$, the triangular inequality and (\ref{weqn18}), we have
\begin{eqnarray}\label{wweqn11}
&  & \|\mathbf{\tilde{X}}\boldsymbol{\bbeta^{*}} - \mathbf{\tilde{X}}\hat{\bbeta}(\mbox{AdaGril})\|_{2}^{2} + \frac{\lambda_1^{*}}{2}\|\boldsymbol{\bbeta^{*}} - \hat{\bbeta}(\mbox{AdaGril})\|_{1} \nonumber\\
& \leq & \lambda_1^{*}\sum_{j \in \mathcal{A}} \hat{\omega}_{j}(|\bbeta^{*}_{j}| - |\hat{\bbeta}(\mbox{AdaGril})_{j}| + |\bbeta^{*}_{j} - \hat{\bbeta}(\mbox{AdaGril})_{j}|\nonumber\\
& \leq & 2\lambda_1^{*}\sum_{j \in \mathcal{A}} \hat{\omega}_{j}|\bbeta^{*}_{j} - \hat{\bbeta}(\mbox{AdaGril})_{j}|\nonumber\\
& \leq & 2\lambda_1^{*} (\omega_{\max}(\mathcal{A}))\sum_{j \in \mathcal{A}}|\bbeta^{*}_{j} - \hat{\bbeta}(\mbox{AdaGril})_{j}|\nonumber\\
& \leq & 2\lambda_1^{*} \max\left\{\frac{2}{\eta}, 1\right\}\sum_{j \in \mathcal{A}}|\bbeta^{*}_{j} - \hat{\bbeta}(\mbox{AdaGril})_{j}|
\end{eqnarray}
Since
$\sum_{j \in \mathcal{A}} |\bbeta^{*}_{j} - \hat{\bbeta}(\mbox{AdaGril})_{j}|
 \leq  \sqrt{s}\|\bbeta_{\mathcal{A}}^{*} - \hat{\bbeta}(\mbox{AdaGril})_{\mathcal{A}}\|_{2},$
we obtain that
\begin{eqnarray}\label{wweqn13}
\|\mathbf{\tilde{X}}\boldsymbol{\bbeta^{*}} - \mathbf{\tilde{X}}\hat{\bbeta}(\mbox{AdaGril})\|_{2}^{2} + \frac{\lambda_1^{*}}{2}\|\boldsymbol{\bbeta^{*}} - \hat{\bbeta}(\mbox{AdaGril})\|_{1}
& \leq & 2 \sqrt{s} \lambda_1^{*} \max\left\{\frac{2}{\eta}, 1\right\}\|\bbeta_{\mathcal{A}}^{*} - \hat{\bbeta}(\mbox{AdaGril})_{\mathcal{A}}\|_{2}.\nonumber\\
\end{eqnarray}
According to inequality (\ref{wweqn11}), we have
\begin{eqnarray}\label{wweqn14}
\|\boldsymbol{\bbeta^{*}} - \hat{\bbeta}(\mbox{AdaGril})\|_{1}
& \leq & 4\max\left\{\frac{2}{\eta}, 1\right\} \sum_{j \in \mathcal{A}}|\bbeta^{*}_{j} - \hat{\bbeta}(\mbox{AdaGril})_{j}|.
\end{eqnarray}
So
\begin{eqnarray}\label{wweqn15}
\sum_{j \in \mathcal{A}^{c}}|\bbeta^{*}_{j} - \hat{\bbeta}(\mbox{AdaGril})_{j}|)
& \leq & 4 \max\left\{\frac{2}{\eta}, 1\right\} \sum_{j \in \mathcal{A}} |\bbeta^{*}_{j} - \hat{\bbeta}(\mbox{AdaGril})_{j}|.
\end{eqnarray}
This last inequality shows that $\boldsymbol{\bbeta^{*}} - \hat{\bbeta}(\mbox{AdaGril})$ obeys to the assumption RE, and hence
\begin{eqnarray}\label{wweqn16}
\|\bbeta_{\mathcal{A}}^{*} - \hat{\bbeta}(\mbox{AdaGril})_{\mathcal{A}}\|_{2}^{2}
& \leq & \|\bbeta^{*} - \hat{\bbeta}(\mbox{AdaGril})\|_{2}^{2}\nonumber\\
& \leq & \psi^{-1}\|\mathbf{\tilde{X}}\boldsymbol{\bbeta^{*}} - \mathbf{\tilde{X}}\hat{\bbeta}(\mbox{AdaGril})\|_{2}^{2}.
\end{eqnarray}
The combination  of this last inequality with (\ref{wweqn13}), give us
\begin{eqnarray}\label{wweqn17}
\|\mathbf{\tilde{X}}\boldsymbol{\bbeta^{*}} - \mathbf{\tilde{X}}\hat{\bbeta}(\mbox{AdaGril})\|_{2}^{2} + \frac{\lambda_1^{*}}{2}\|\boldsymbol{\bbeta^{*}} - \hat{\bbeta}(\mbox{AdaGril})\|_{1}
& \leq & 2\sqrt{s}\lambda_1^{*} \sqrt{\psi^{-1}}\max\left\{\frac{2}{\eta}, 1\right\}\times\nonumber\\
&   &  \|\mathbf{\tilde{X}}\boldsymbol{\bbeta^{*}} - \mathbf{\tilde{X}}\hat{\bbeta}(\mbox{AdaGril})\|_{2}
\end{eqnarray}
So
\begin{eqnarray}\label{wweqn18}
\|\mathbf{\tilde{X}}\boldsymbol{\bbeta^{*}} - \mathbf{\tilde{X}}\hat{\bbeta}(\mbox{AdaGril})\|_{2}^{2}
& \leq & 4\psi^{-1}\lambda_1^{*2}\left(\max\left\{\frac{2}{\eta}, 1\right\}\right)^{2} s.
\end{eqnarray}
Since
\begin{eqnarray}\label{wweqn20}
\|\mathbf{\tilde{X}}\boldsymbol{\bbeta^{*}} - \mathbf{\tilde{X}}\hat{\bbeta}(\mbox{AdaGril})\|_{2}^{2}
& = & \|\mathbf{X}\boldsymbol{\bbeta^{*}} - \mathbf{X}\hat{\bbeta}(\mbox{AdaGril})\|_{2}^{2}\nonumber\\
& + & \lambda_{2}(\bbeta^{*} - \hat{\bbeta}(\mbox{AdaGril}))^t\mathbf{Q}(\bbeta^{*} - \hat{\bbeta}(\mbox{AdaGril})),
\end{eqnarray}
we obtain
\begin{eqnarray}\label{wweqn21}
\|\mathbf{X}\boldsymbol{\bbeta^{*}} - \mathbf{X}\hat{\bbeta}(\mbox{AdaGril})\|_{2}^{2}
& \leq & 4\psi^{-1}\lambda_1^{*2}\left(\max\left\{\frac{2}{\eta}, 1\right\}\right)^{2} s.
\end{eqnarray}
Using (\ref{wweqn13}) and the fact that $\|\mathbf{v}\|_{\infty} \leq \|\mathbf{v}\|_{1}$ for all $\mathbf{v} \in \mathbb{R}^{p}$, we have
\begin{eqnarray}\label{wweqn22}
\|\boldsymbol{\bbeta^{*}} - \hat{\bbeta}(\mbox{AdaGril})\|_{1}
& \leq & 8\psi^{-1}\lambda_1^{*}\left(\max\left\{\frac{2}{\eta}, 1\right\}\right)^{2} s,
\end{eqnarray}
and
\begin{eqnarray}\label{wweqn23}
\|\boldsymbol{\bbeta^{*}} - \hat{\bbeta}(\mbox{AdaGril})\|_{\infty}
& \leq & 8\psi^{-1}\lambda_1^{*}\left(\max\left\{\frac{2}{\eta}, 1\right\}\right)^{2} s.
\end{eqnarray}
\end{proof}
\begin{center}
{\bf Acknowledgements}
\end{center}
We warmly thank the reviewer for his (or her)
careful reading of the previous version of our paper and helpful
comments. The first author is partially supported by the Maroc-Stic program.

\begin{table}[htbp]
  \begin{center}
\begin{tabular}{lcccccccc}
 \hline \hline
  & \multicolumn{2}{c}{ $\sigma$ = 3 } && \multicolumn{2}{c}{ $\sigma$ = 6 } && \multicolumn{2}{c}{ $\sigma$ = 9 } \\ \cline{2-3} \cline{5-6} \cline{8-9}
                             & $\rho$ = 0.5 & $\rho$ = 0.75 && $\rho$ = 0.5 & $\rho$ = 0.75 && $\rho$ = 0.5 & $\rho$ = 0.75 \\
                  \cline{2-3} \cline{5-6} \cline{8-9}
 $\mathbf{n=100}$                &&&&&&&&\\
 Lasso                        & 2.51 & 2.30   && 10.52 & 10.47 && 24.50 & 19.53 \\
 AdaLasso                     & 1.74 & 1.73   && 8.38  & 9.61  && 22.95 & 19.26 \\
 Enet                         & 2.32 & 2.14   && 9.52  & 8.88  && 19.89 & 17.52 \\
 AdaEnet                      & 1.89 & 1.83   && 8.90  & 8.67  && 21.71 & 18.58 \\
 Slasso                       & 2.42 & 2.21   && 9.70  & 9.21  && 20.14 & 16.75 \\
 AdaSlasso                    & 1.60 & 1.44   && 7.67  & 8.30  && 20.25 & 16.76 \\
 Cnet                         & 2.43 & 2.16   && 9.67  & 9.03  && 19.89 & 17.42 \\
 AdaCnet                      & 1.81 & 1.72   && 8.64  & 8.57  && 21.24 & 17.92 \\
 Wfusion                      & 2.35 & 2.25   && 9.52  & 8.67  && 19.93 & 17.55 \\
 AdaWfusion                   & 1.89 & 1.94   && 8.80  & 8.77  && 21.29 & 18.30 \\  \cline{1-9}
 $\mathbf{n=200}$                &&&&&&&&\\
 Lasso                        & 1.74 & 1.62   && 7.91 & 7.26 && 16.31 & 15.02 \\
 AdaLasso                     & 1.04 & 0.99   && 5.06 & 5.54 && 13.04 & 14.80 \\
 Enet                         & 1.62 & 1.54   && 7.23 & 6.67 && 14.63 & 13.92 \\
 AdaEnet                      & 1.17 & 1.05   && 5.35 & 5.53 && 13.02 & 13.91 \\
 Slasso                       & 1.89 & 1.77   && 7.70 & 6.63 && 14.82 & 13.67 \\
 AdaSlasso                    & 0.99 & 0.94   && 4.71 & 4.77 && 11.40 & 11.94 \\
 Cnet                         & 1.93 & 1.80   && 7.75 & 6.91 && 15.09 & 13.92 \\
 AdaCnet                      & 1.08 & 0.98   && 5.12 & 5.41 && 12.47 & 13.54 \\
 Wfusion                      & 1.63 & 1.54   && 7.23 & 6.68 && 14.63 & 13.92 \\
 AdaWfusion                   & 1.10 & 1.01   && 5.25 & 5.43 && 12.76 & 13.74 \\  \cline{1-9}
 $\mathbf{n=1000}$                &&&&&&&&\\
 Lasso                        & 0.84 & 0.72   && 3.50 & 2.79 && 7.62 & 6.55 \\
 AdaLasso                     & 0.59 & 0.50   && 2.28 & 2.21 && 5.30 & 4.85 \\
 Enet                         & 0.82 & 0.68   && 3.33 & 2.72 && 7.27 & 6.39 \\
 AdaEnet                      & 1.03 & 1.29   && 2.80 & 3.08 && 6.12 & 5.82 \\
 Slasso                       & 1.32 & 1.35   && 3.89 & 3.37 && 8.04 & 7.10 \\
 AdaSlasso                    & 0.73 & 0.80   && 2.37 & 2.29 && 5.19 & 4.66 \\
 Cnet                         & 3.32 & 2.71   && 6.83 & 5.54 && 12.0 & 9.81 \\
 AdaCnet                      & 0.41 & 0.40   && 1.92 & 1.89 && 4.57 & 4.92 \\
 Wfusion                      & 0.82 & 0.68   && 3.33 & 2.72 && 7.27 & 6.39 \\
 AdaWfusion                   & 0.60 & 0.53   && 2.39 & 2.24 && 5.55 & 4.93 \\  \cline{1-9}
 \hline\hline
\\ [-.15in]
\\
\end{tabular}
\caption{Median mean-squared errors for $\rho \in \{0.5, 0.75\}$,  $\sigma \in \{3, 6, 9\}$ and $n \in \{100, 200, 1000\}$ based on $100$ replications.}\label{gril.tab.1}
\end{center}
\end{table}

\begin{table}[htbp]
  \begin{center}
\begin{tabular}{lcccccccc}
 \hline \hline
  & \multicolumn{2}{c}{ $\sigma$ = 3 } && \multicolumn{2}{c}{ $\sigma$ = 6 } && \multicolumn{2}{c}{ $\sigma$ = 9 } \\ \cline{2-3} \cline{5-6} \cline{8-9}
                             & $\rho$ = 0.5 & $\rho$ = 0.75 && $\rho$ = 0.5 & $\rho$ = 0.75 && $\rho$ = 0.5 & $\rho$ = 0.75 \\
                  \cline{2-3} \cline{5-6} \cline{8-9}
 $\mathbf{n=100}$                &&&&&&&&\\
 Lasso                        & 3.30 & 6.51   && 12.66 & 22.70 && 26.41 & 32.97 \\
 AdaLasso                     & 2.64 & 5.57   && 13.65 & 32.91 && 35.88 & 52.01 \\
 Enet                         & 3.12 & 6.02   && 12.08 & 20.51 && 23.92 & 33.03 \\
 AdaEnet                      & 2.75 & 5.44   && 14.35 & 28.79 && 34.10 & 52.08 \\
 Slasso                       & 3.24 & 6.21   && 11.82 & 20.13 && 23.23 & 30.14 \\
 AdaSlasso                    & 2.30 & 4.14   && 11.83 & 26.22 && 30.74 & 45.88 \\
 Cnet      & 3.20 & 6.03   && 12.10 & 20.62 && 23.61 & 31.72 \\
 AdaCnet   & 2.62 & 5.15   && 13.87 & 28.43 && 33.15 & 50.33 \\
 Wfusion                      & 3.18 & 6.46   && 12.08 & 20.52 && 23.93 & 33.01 \\
 AdaWfusion                   & 2.75 & 5.92   && 14.12 & 28.38 && 33.49 & 51.26 \\  \cline{1-9}
 $\mathbf{n=200}$                &&&&&&&&\\
 Lasso                        & 2.27 & 4.53   && 9.40 & 19.00 && 20.27 & 35.36 \\
 AdaLasso                     & 1.54 & 3.05   && 7.56 & 19.17 && 21.95 & 53.93 \\
 Enet                         & 2.15 & 4.29   && 8.79 & 17.58 && 18.70 & 33.77 \\
 AdaEnet                      & 1.65 & 3.06   && 7.67 & 17.97 && 20.29 & 47.83 \\
 Slasso                       & 2.56 & 5.18   && 9.12 & 16.61 && 18.25 & 31.00 \\
 AdaSlasso                    & 1.46 & 2.77   && 6.61 & 14.44 && 17.47 & 38.40 \\
 Cnet      & 2.45 & 4.89   && 9.14 & 17.57 && 18.81 & 32.66 \\
 AdaCnet   & 1.58 & 2.90   && 7.30 & 17.59 && 19.76 & 46.73 \\
 Wfusion                      & 2.16 & 4.29   && 8.79 & 17.59 && 18.70 & 33.77 \\
 AdaWfusion                   & 1.60 & 2.99   && 7.52 & 17.67 && 19.92 & 47.07 \\  \cline{1-9}
 $\mathbf{n=1000}$                &&&&&&&&\\
 Lasso                        & 0.95 & 1.95   && 3.88 & 7.77  && 8.39  & 17.66 \\
 AdaLasso                     & 0.83 & 1.37   && 3.25 & 6.05  && 7.34  & 13.63 \\
 Enet                         & 0.94 & 1.88   && 3.77 & 7.62  && 8.16  & 17.35 \\
 AdaEnet                      & 1.01 & 1.59   && 3.58 & 6.38  && 7.93  & 14.06 \\
 Slasso                       & 1.81 & 4.94   && 4.69 & 10.35 && 9.10  & 19.32 \\
 AdaSlasso                    & 1.23 & 3.03   && 3.46 & 6.52  && 7.09  & 12.86 \\
 Cnet      & 2.74 & 6.04   && 6.10 & 12.69 && 11.14 & 22.12 \\
 AdaCnet   & 0.64 & 1.29   && 2.91 & 5.69  && 6.64  & 16.02 \\
 Wfusion                      & 0.94 & 1.88   && 3.77 & 7.62  && 8.16  & 17.36 \\
 AdaWfusion                   & 0.84 & 1.40   && 3.37 & 6.09  && 7.62  & 13.65 \\  \cline{1-9}
 \hline\hline
\\ [-.15in]
\\
\end{tabular}
\caption{$\mbox{MSE}_{\beta} = \|\hat{\beta} - \beta^{*}\|_{2}^{2}$ errors for $\rho \in \{0.5, 0.75\}$,  $\sigma \in \{3, 6, 9\}$ and $n \in \{100, 200, 1000\}$ based on $100$ replications.}\label{gril.tab.2}
\end{center}
\end{table}

    \begin{table}[htbp]
  \begin{center}
\begin{tabular}{lcccccccc}
 \hline \hline
  & \multicolumn{2}{c}{ $\sigma$ = 3 } && \multicolumn{2}{c}{ $\sigma$ = 6 } && \multicolumn{2}{c}{ $\sigma$ = 9 } \\ \cline{2-3} \cline{5-6} \cline{8-9}
                             & C & IC && C & IC && C & IC \\
                  \cline{2-3} \cline{5-6} \cline{8-9}
 $\mathbf{n=100}$                &&&&&&&&\\
 Lasso                        & 22.23 & 0.11   && 23.10 & 1.16 && 24.35 & 3.38 \\
 AdaLasso                     & 25.06 & 0.53   && 25.20 & 2.03 && 25.39 & 4.45 \\
 Enet                         & 20.83 & 0.09   && 21.45 & 0.98 && 22.53 & 2.47 \\
 AdaEnet                      & 24.37 & 0.31   && 24.47 & 1.72 && 24.67 & 3.65 \\
 Slasso                       & 21.45 & 0.06   && 21.90 & 0.93 && 22.99 & 2.48 \\
 AdaSlasso                    & 24.76 & 0.29   && 24.66 & 1.57 && 24.79 & 3.55 \\
 Cnet                         & 21.04 & 0.08   && 21.51 & 0.96 && 22.67 & 2.47 \\
 AdaCnet                      & 24.49 & 0.31   && 24.50 & 1.72 && 24.75 & 3.67 \\
 Wfusion                      & 20.64 & 0.09   && 21.45 & 0.98 && 22.54 & 2.47 \\
 AdaWfusion                   & 24.13 & 0.31   && 24.46 & 1.71 && 24.68 & 3.65 \\  \cline{1-9}
 $\mathbf{n=200}$                &&&&&&&&\\
 Lasso                        & 31.64 & 0.02   && 32.40 & 0.43 && 32.47 & 1.12 \\
 AdaLasso                     & 35.12 & 0.07   && 35.19 & 1.23 && 35.39 & 2.59 \\
 Enet                         & 30.77 & 0.02   && 31.21 & 0.37 && 31.08 & 0.87 \\
 AdaEnet                      & 34.67 & 0.05   && 34.58 & 1.00 && 34.68 & 1.98 \\
 Slasso                       & 31.26 & 0.02   && 31.75 & 0.34 && 31.37 & 0.85 \\
 AdaSlasso                    & 35.08 & 0.05   && 34.75 & 0.94 && 34.83 & 1.87 \\
 Cnet                         & 31.67 & 0.02   && 31.66 & 0.37 && 31.27 & 0.86 \\
 AdaCnet                      & 34.80 & 0.05   && 34.65 & 0.97 && 34.78 & 2.02 \\
 Wfusion                      & 30.87 & 0.02   && 31.21 & 0.37 && 31.08 & 0.87 \\
 AdaWfusion                   & 34.67 & 0.05   && 34.58 & 1.00 && 34.68 & 1.98 \\  \cline{1-9}
 $\mathbf{n=1000}$                &&&&&&&&\\
 Lasso                        & 76.16 & 0.00   && 76.63 & 0.00 && 76.54 & 0.05 \\
 AdaLasso                     & 76.16 & 0.00   && 76.63 & 0.00 && 76.54 & 0.05 \\
 Enet                         & 75.76 & 0.00   && 75.67 & 0.00 && 75.49 & 0.04 \\
 AdaEnet                      & 75.77 & 0.00   && 75.67 & 0.00 && 75.49 & 0.04 \\
 Slasso                       & 77.13 & 0.00   && 76.09 & 0.00 && 76.15 & 0.05 \\
 AdaSlasso                    & 77.13 & 0.00   && 76.09 & 0.00 && 78.76 & 0.07 \\
 Cnet      & 80.64 & 0.00   && 79.15 & 0.00 && 78.76 & 0.07 \\
 AdaCnet   & 80.64 & 0.00   && 79.15 & 0.00 && 78.76 & 0.07 \\
 Wfusion                      & 75.76 & 0.00   && 75.67 & 0.00 && 75.49 & 0.04 \\
 AdaWfusion                   & 75.76 & 0.00   && 75.67 & 0.00 && 75.49 & 0.04 \\  \cline{1-9}
 \hline\hline
\\ [-.15in]
\\
\end{tabular}
\caption{The median number of C and IC $\rho = 0.5$,  $\sigma \in \{3, 6, 9\}$ and $n \in \{100, 200, 1000\}$ based on $100$ replications.}\label{gril.tab.3}
\end{center}
\end{table}
\end{document}